\documentclass[11pt,a4paper]{article}

\usepackage{a4wide}

\usepackage{amsfonts}
\usepackage{amssymb}
\usepackage{graphicx,color}
\usepackage{amsmath}

\usepackage{bm}
\usepackage{pifont}

\usepackage{fontawesome}
\usepackage{lscape}
\usepackage{pdflscape}
\usepackage{transparent}

\usepackage[notref,notcite]{showkeys}

\usepackage{appendix}
\newtheorem{remark}{Remark}

\usepackage{authblk}
\date{\today}
\author[1]{Tymofiy Gerasimov}
\author[2]{Laura De Lorenzis}
\affil[1]{Institute of Applied Mechanics,\authorcr Technische Universit\"at Braunschweig, Germany}
\affil[2]{Department of Mechanical and Process Engineering,\authorcr ETH Z\"urich, Switzerland}

\usepackage{lipsum}

\newbox\flinebox 
\newbox\slinebox
\newbox\mlinebox
\def\duplines{\setlength\parindent{0pt}
  \setbox\flinebox\lastbox
  \ifvoid\flinebox\relax
  \else
  \setbox\slinebox\hbox{\copy\flinebox}
  \setbox\mlinebox\hbox{\copy\flinebox}
  \unskip\unpenalty
  {\duplines}
  \box\flinebox\vspace*{-2.3ex}
  \box\mlinebox\vspace*{-2.3ex}
  \box\slinebox \fi
}

\begin{document}
%\transparent{0.3}

\title{Second-order phase-field formulations for anisotropic brittle fracture}

\maketitle

\begin{abstract}
We address brittle fracture in anisotropic materials featuring two-fold and four-fold symmetric fracture toughness. For these two classes, we develop two variational phase-field models based on the family of regularizations proposed by Focardi (Focardi, M. On the variational approximation of free-discontinuity problems in the vectorial case.\
Math.\ Models Methods App.\ Sci., 11:663--684, 2001), for which $\Gamma$-convergence results hold. Since both models are of second order, as opposed to the previously available fourth-order models for four-fold symmetric fracture toughness, they do not require  basis functions of $C^1$-continuity nor mixed variational principles for finite element discretization. For the four-fold symmetric formulation we show that the standard quadratic degradation function is unsuitable and devise a procedure to derive a suitable one. The performance of the new models is assessed via several numerical examples that simulate anisotropic fracture under anti-plane shear loading. For both formulations at fixed displacements (i.e. within an alternate minimization procedure), we also provide some existence and uniqueness results for the phase-field solution.
\end{abstract}

%$C^1$ finite element. For this formulation we show that the choice of a degradation (also termed coupling) function is not straightforward, and that the correct one calls for the necessity of having a robust solution procedure that can cope with non-convexity of the corresponding minimization problem. To this end, we propose using the trust-region method. Finally, our models'

\vspace{0.2cm}
\vspace{0.5cm}

\newpage
\tableofcontents

%%%%%%%%%%%%%%%%%%%%%%%%%%%%%%%%%%%%%%%%%%%%%%%%%%%%%%%%
%\section{Introduction}
%{\color{red}...LDL...}

%%%%%%%%%%%%%%%%%%%%%%%%%%%%%%%%%%%%%%%%%%%%%%%%%%%%%%%%
\section{Introduction}%\section{Anisotropic fracture surface energy}

Predicting the fracture behavior in brittle materials with anisotropic fracture toughness is still a challenge. The extension to the anisotropic case of selection criteria for the crack direction which have proved successful in fracture mechanics of isotropic materials, such as the principle of local symmetry \cite{Goldstein,Cotterell} and the maximum energy release rate criterion \cite{Hussain}, is by no means trivial \cite{Marder}. The generalized energy release rate criterion \cite{Palaniswamy,Gurtin,Marder}, where the crack propagation direction is chosen by maximizing the ratio of the energy release rate to the fracture toughness with respect to the propagation angle, was shown to correlate successfully with experiments on materials exhibiting both weak \cite{Ibarra} and strong \cite{Takei} anisotropy in the fracture toughness, see also the detailed review and discussion in \cite{Li2019}. 

The variational phase-field approach to fracture, pioneered by Bourdin et al. \cite{Bourdin2000} and first proposed as the regularization of Francfort and Marigo's variational fracture formulation \cite{Francfort1998}, was recently extended to materials with anisotropic fracture toughness. Hakim and Karma \cite{Hakim2005, Hakim2009} proposed a phase-field model for materials featuring two-fold weak anisotropy and elucidated several features of crack propagation in these materials. This model was extended to the three-dimensional setting including geometric nonlinearity and nonlinear elasticity \cite{Clayton} and to the consideration of multiple cleavage planes using multiple phase-field variables \cite{Nguyen}. With the purpose of tackling phenomena such as sawtooth crack patterns and forbidden crack directions, which are directly related to four-fold strongly anisotropic surface energies, Li et al. \cite{Li2015} formulated a fourth-order anisotropic variational fracture model inspired by phase-field models of crystal growth. Similarly, Teichtmeister et al. \cite{Teichtmeister} formulated second- and fourth-order phase-field models based on structural tensors for two-fold and four-fold symmetry, respectively, and extended the framework to large deformations. Li and Maurini \cite{Li2019} improved and simplified the phase-field model in \cite{Li2015} in order to enable an analytical solution for the optimal crack profile, while keeping the fundamental properties of the phase-field approximation. Once again, four-fold symmetry was addressed with a fourth-order model.

In this paper, we address brittle fracture in anisotropic materials featuring two-fold and four-fold symmetric fracture toughness and develop two variational phase-field models based on the family of regularizations proposed by Focardi \cite{Focardi2001}, who also proved their $\Gamma$-convergence. The main novelty lies in the model for four-fold symmetric anisotropy: as opposed to all previously available models for the same type of anisotropy, our proposed model is of second order, hence it does not require higher-continuity basis functions nor mixed variational principles for finite element discretization. For the four-fold symmetric formulation we show that the standard quadratic degradation function is unsuitable and propose a procedure to derive a suitable one. The performance of the new models is assessed via several numerical examples that simulate anisotropic fracture under anti-plane shear loading. For both formulations at fixed displacements, we also provide some existence and uniqueness results for the phase-field solution.

The remainder of this manuscript is organized as follows. Section \ref{sec:probsetting} introduces basic concepts on anisotropy of fracture properties that serve as basis for the following investigation. In Section \ref{sec:varform}, we  recall the phase-field formulation of (isotropic) brittle fracture based on the Ambrosio-Tortorelli regularization and introduce the main results in \cite{Focardi2001}, which are our point of departure for the new developments in the subsequent sections. We show that the models based on the Focardi regularization combined with the Euclidean norm, which we term isotropic \texttt{Foc-}$p$ models, can be viewed as a family of models which includes the well-known \texttt{AT-}2 model as a special case (corresponding to the isotropic \texttt{Foc-}$2$ model). In Section \ref{Res1}, we focus on the isotropic \texttt{Foc-}$4$ model. We show that the standard quadratic degradation function is unsuitable for this model, we derive a new suitable one, and we test the ensuing formulation on an anti-plane shear numerical setup. In Section  \ref{Res2}, we develop the anisotropic counterparts of the isotropic \texttt{Foc-}2 and \texttt{Foc-}4 formulations, which model an anisotropic fracture behavior with two- and four-fold symmetric fracture toughness, respectively. Section \ref{NumTest} illustrates the numerical tests on the two anisotropic formulations for different anti-plane shear setups. We draw the main conclusions in Section \ref{Conclusions}. Appendix \ref{Ap1} reports some existence and uniqueness results on both anisotropic formulations. 
%whereas Appendix \ref{Ap2} attempts an explanation of the micro zig-zagging phenomenon observed in the numerical experiments of Section \ref{Res2}.

\section{Basic concepts on anisotropic fracture toughness}%\section{Anisotropic fracture surface energy}
\begin{figure}[!ht]
\begin{center}
\includegraphics[width=1.0\textwidth]{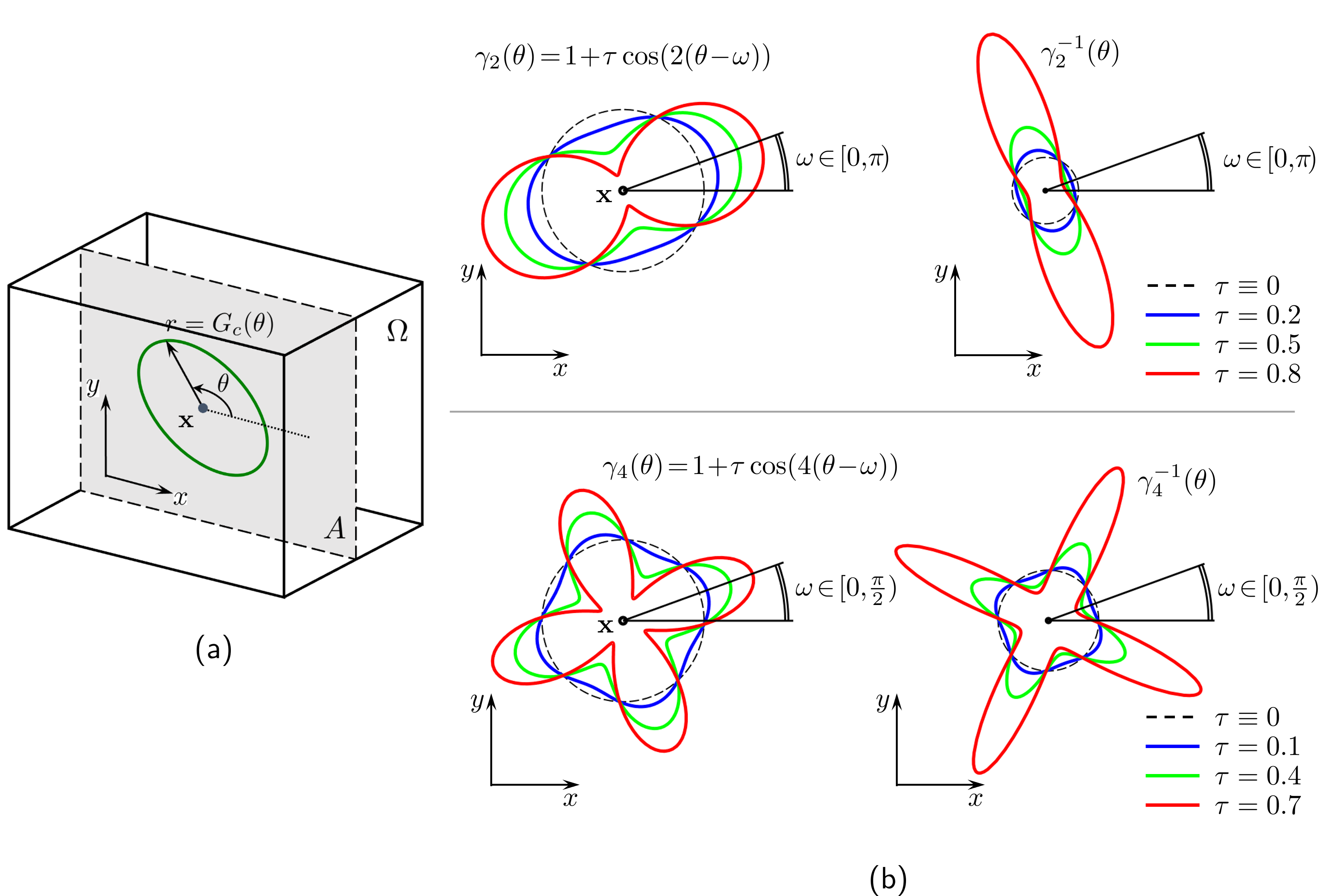}
\end{center}
\caption{Sketch of anisotropy on a plane $A$ (a) and plots of $\gamma_2$, $\gamma_2^{-1}$, and $\gamma_4$, $\gamma_4^{-1}$ in $A$ for representing two- and four-fold symmetric anisotropic fracture toughness, respectively (b).}
\label{Fig:1}
\end{figure}

\label{sec:probsetting}
Materials with anisotropic fracture behavior are characterized by an orientation-dependent fracture toughness or critical energy release rate $G_c$. One of the examples is sketched in Figure \ref{Fig:1}a, where it is assumed that there exists a plane of anisotropy $A\subset\mathbb{R}^2$ within the body $\Omega\subset\mathbb{R}^3$ such that for any point ${\bf x}\in A$ and with the polar coordinate system within $A$ associated to $\bf x$, $G_c$ is viewed as a function of the polar angle $\theta\in[0,2\pi)$, that is, $G_c=G_c(\theta)$. This is in contrast to the simple isotropic case when $G_c\equiv\mathrm{const}$.
In this paper, we consider the two-dimensional setting by taking $\Omega\subset\mathbb{R}^2$, which implies that the anisotropy plane $A$ and $\Omega$ coincide.
For modeling purposes, it is furthermore assumed that 
\begin{equation}
G_c(\theta)=G_0\gamma(\theta),
\label{GcTheta}
\end{equation}
where $G_0$ is a positive constant and $\gamma(\theta)$ is a periodic trigonometric function. A suitable form of this function reads
\begin{equation}
\gamma(\theta)=\gamma_k(\theta):=1+\tau\cos(k(\theta-\omega)),
\label{gammaK}
\end{equation}
with $k\in\mathbb{N}$, $k\geq2$, $\tau\in[0,1)$ and $\omega\in[0,\frac{2\pi}{k})$. The parameter $\omega$ defines the direction corresponding to the largest fracture toughness and hence it is referred to as the {\em strongest} or {\em principal} material direction. The parameter $\tau$ defines the magnitude of the largest fracture toughness and is called the {\em anisotropy strength}. A useful induced quantity for characterization of the anisotropy strength is the ratio $\frac{1+\tau}{1-\tau}\in[1,\infty)$. Finally, $k$ is the periodicity parameter such that the function $\gamma_k$ is $\frac{2\pi}{k}$-periodic. 
The above gives rise to the notion of a $k$-{\em fold} symmetric fracture toughness $G_c$.

In the present work, we are interested in the cases $k=2$ and $4$ in (\ref{gammaK}), i.e. in the so-called two- and four-fold anisotropy cases. We give the corresponding explicit expressions of $\gamma_k$ for future reference:

\begin{align}
\gamma_2(\theta)=&1+\tau\cos(2(\theta-\omega)) \nonumber \\
    =&1+\tau
    \left[ 
    \cos(2\omega)(\cos^2(\theta)-\sin^2(\theta))
    +2\sin(2\omega)\sin(\theta)\cos(\theta)
    \right],
\label{gamma2}
\end{align}
with $\tau\in[0,1)$, $\omega\in[0,\pi)$, and
\begin{align}
\gamma_4(\theta)=1+&\tau\cos(4(\theta-\omega)) \nonumber \\
    =1+&\tau
    \left[\cos(4\omega)(\cos^4(\theta)-6\sin^2(\theta)\cos^2(\theta)+\sin^4(\theta)) 
    \right. \nonumber \\
    &+
    \left.
    4\sin(4\omega)\sin(\theta)\cos(\theta)(\cos^2(\theta)-\sin^2(\theta))
    \right],
\label{gamma4}
\end{align} 
with $\tau\in[0,1)$, $\omega\in[0,\frac{\pi}{2})$.
The polar plots of both functions are continuous closed curves surrounding
the origin of the polar coordinate system. 
According to standard terminology, the anisotropy is said to be \textit{weak}
if $\gamma\left(\theta\right)$ is convex and \textit{strong} if $\gamma\left(\theta\right)$
is non-convex. The convexity of the function $\gamma\left(\theta\right)$
is known to be equivalent to the convexity of the region enclosed
by the polar plot of $\gamma^{-1}\left(\theta\right)$, which is guaranteed
if

\begin{equation}
\gamma\left(\theta\right)+\gamma''\left(\theta\right)\geq0\qquad\forall\,\theta\in\left[0,2\pi\right)
\end{equation}
With the expressions of $\gamma_{2}\left(\theta\right)$ and $\gamma_{4}\left(\theta\right)$
in (\ref{gamma2}) and (\ref{gamma4}), it is straightforward to verify that

\begin{equation}
\label{convexity2}
\gamma_{2}\left(\theta\right)\:\mathrm{convex}\quad\Leftrightarrow\quad\tau\in\left[0,\frac{1}{3}\right]
\end{equation}
and

\begin{equation}
\label{convexity4}
\gamma_{4}\left(\theta\right)\:\mathrm{convex}\quad\Leftrightarrow\quad\tau\in\left[0,\frac{1}{15}\right]
\end{equation}
regardless of the value of $\omega$. The polar plots of $\gamma_{2}\left(\theta\right)$ and  $\gamma_{2}^{-1}\left(\theta\right)$, as well as $\gamma_{4}\left(\theta\right)$ and $\gamma_{4}^{-1}\left(\theta\right)$
for a fixed $\omega$ and different magnitudes of $\tau\in[0,1)$ are depicted in Figure \ref{Fig:1}b. With the increase of $\tau$, the elliptic shape of $\gamma_2$ and $\gamma_2^{-1}$ turns into a peanut-like shape, whereas for $\gamma_4$ and $\gamma_4^{-1}$ one has a transformation from super-ellipse to 'butterfly'.

In the following, when considering the cases of isotropic and anisotropic fracture toughness we refer to them as 'isotropic fracture' and 'anisotropic fracture', respectively. In the former case, we assume $G_c(\theta)\equiv\mathrm{const}$ such that, owing to (\ref{GcTheta}) where $\gamma(\theta)\equiv 1$, it is simply $G_c(\theta)=G_0$. In the latter case, the term 'anisotropic fracture' is accompanied by the specification 'with $k$-fold symmetric $G_c$'.

%%%%%%%%%%%%%%%%%%%%%%%%%%%%%%%%%%%%%%%%%%%%%%%%%%%%%%%%
\section{Variational formulation of brittle fracture and phase-field regularization}
\label{sec:varform}
In this section, we briefly recall the phase-field formulation of (isotropic) brittle fracture from \cite{Bourdin2000,Bourdin2007a,Bourdin2007b,Bourdin2008}, based on the so-called Ambrosio-Tortorelli regularization \cite{AT1992,AT1990}. This is followed by the presentation of the main results by Focardi \cite{Focardi2001}, which are our point of departure for the new developments in the subsequent sections. Finally, we briefly outline the incremental minimization problem which is solved for computing a quasi-static phase-field evolution.

\begin{figure}[!ht]
\begin{center}
\includegraphics[width=1.0\textwidth]{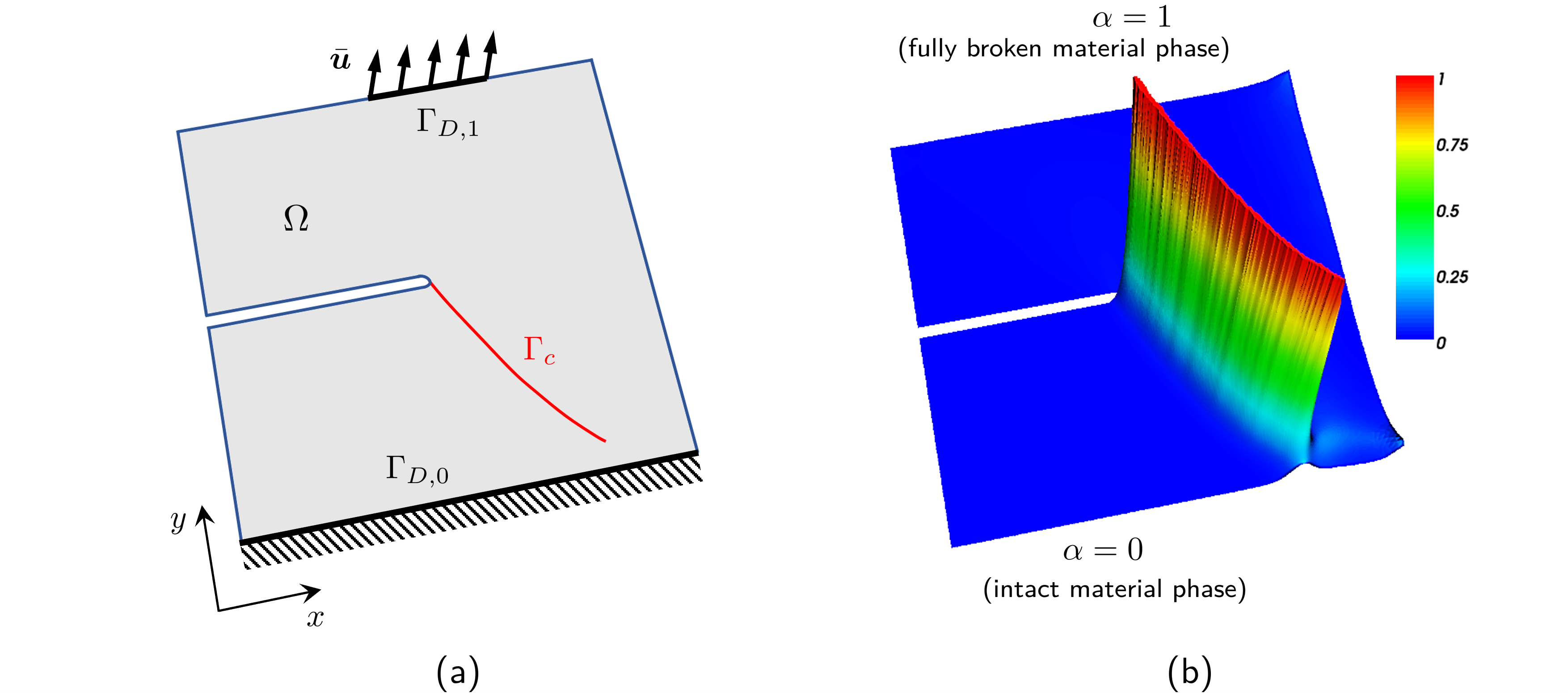}
\end{center}
\caption{Two-dimensional body with a sharp crack (a) and with the same crack modeled through a phase-field variable $\alpha\in C(\Omega;[0,1])$ (b).}
\label{Fig:4}
\end{figure}

Let $\Omega\subset\mathbb{R}^d$, $d=2$ or $3$ be an open and bounded domain representing the configuration of a $d$-dimensional body, and let $\Gamma_{D,0}$ and $\Gamma_{D,1}$ be the (non-overlapping) portions of the boundary $\partial\Omega$ of $\Omega$ which are fixed and subjected to the imposed displacement $\bar{\bm u}$, respectively (Figure \ref{Fig:4}a). Let also  $\Gamma_c\subset\Omega$ represent the crack set, that is, the set of points where the displacement field $\bm u$ is discontinuous.

\subsection{Ambrosio-Tortorelli regularization}
\label{AT_reg}
The variational approach to (isotropic) brittle fracture by Francfort and Marigo \cite{Francfort1998} relies on the energy functional
\begin{equation}
\mathcal{E}(\bm u,\Gamma_c) = E_\mathrm{el.}(\bm u,\Gamma_c)+E_S(\Gamma_c)
= \int_{\Omega\setminus\Gamma_c} 
\Psi(\bm\varepsilon(\bm u)) \, \mathrm{d}{\bf x}
+ G_0\int_{\Gamma_c} \mathrm{d}\mathcal{H}^{d-1},
\label{VAF}
\end{equation}
and the related incremental minimization problem. In \eqref{VAF}, $\Psi$ is the elastic energy density, a function of the infinitesimal strain $\bm\varepsilon(\bm u) = \tfrac{1}{2}(\nabla\bm u + (\nabla\bm u)^T)$, where $\bm u:\Omega\rightarrow\mathbb{R}^d$ is the displacement field and $\nabla$ is the gradient operator, and $\mathcal{H}^{d-1}$ is the Hausdorff measure. The terms $E_\mathrm{el.}$ and $E_S$ respectively represent the elastic energy stored in the cracked body and the surface energy associated to the crack. %dissipated within the fracture process,respectively. 
Upon integration, the latter reads $E_S(\Gamma_c)=G_0\mathcal{H}^{d-1}(\Gamma_c)$ where, in simple terms, $\mathcal{H}^1(\Gamma_c)$ and $\mathcal{H}^2(\Gamma_c)$ are the length and the surface area of $\Gamma_c$ when $d=2$ and $3$, respectively. Thus the expression of the surface energy associated to the crack is consistent with Griffith's theory \cite{Griffith} with a fracture toughness $G_0$.

The regularization of (\ref{VAF}) {\em \`a la} Ambrosio-Tortorelli \cite{AT1992} developed by Bourdin-Francfort-Marigo \cite{Bourdin2000,Bourdin2007a,Bourdin2007b,Bourdin2008}, which forms the basis for a variety of (isotropic) fracture phase-field formulations, reads as follows:
\begin{equation}
\mathcal{E}(\bm u,\alpha)
=E_\mathrm{el.}(\bm u,\alpha)+E_S(\alpha)
=\int_\Omega
{\sf g}(\alpha)\Psi(\bm\varepsilon(\bm u)) \, \mathrm{d}{\bf x} 
+  \frac{G_0}{c_{\sf w}} \int_\Omega 
\left(\frac{1}{\ell}{\sf w}(\alpha)+\ell|\nabla\alpha|^2\right)\mathrm{d}{\bf x},
\label{RegVAF}
\end{equation}
with $\alpha:\Omega\rightarrow[0,1]$ as the phase-field variable, which takes the value $1$ on $\Gamma_c$, decays smoothly to $0$ in a subset of $\Omega\backslash\Gamma_c$ and vanishes in the rest of the domain, as sketched in Figure \ref{Fig:4}b. With this definition, the limits $\alpha=1$ and $\alpha=0$ represent the fully broken and the intact (undamaged) material phases, respectively, whereas the intermediate range $\alpha\in(0,1)$ mimics the transition zone between them. The function $\sf g$, often denoted as the degradation or the coupling function, is responsible for the material stiffness degradation, the function $\sf w$, often termed the local dissipation function \cite{Pham2011}, defines the (decaying) profile of $\alpha$, whereas the parameter $0<\ell\ll\mathrm{diam}(\Omega)$ controls the thickness of the localization zone of $\alpha$, i.e.\ of the transition zone between the two material states.

The specific choice of the functions $\sf g$ and $\sf w$ in \eqref{RegVAF} establishes the rigorous link between \eqref{VAF} and (\ref{RegVAF}) when $\ell\rightarrow 0$ via the notion of $\Gamma$-convergence, see e.g. \cite{Braides1998,Chambolle2004,AT1990}, also giving a meaning to the induced constant $c_{\sf w}$. Thus, $\sf g$ is a continuous monotonic function that fulfils the properties: ${\sf g}(0)=1$, ${\sf g}(1)=0$, ${\sf g}^\prime(1)=0$ and ${\sf g}^\prime(\alpha)<0$ for $\alpha\in[0,1)$, see e.g. \cite{Pham2011}. The quadratic polynomial ${\sf g}(\alpha):=(1-\alpha)^2$ is the simplest choice. The function $\sf w$ is continuous and monotonic such that ${\sf w}(0)=0$, ${\sf w}(1)=1$ and ${\sf w}^\prime(\alpha)\geq 0$ for $\alpha\in[0,1]$. The constant $c_{\sf w}:=4\int_0^1\sqrt{{\sf w}(t)}\,\mathrm{d}t$ is a normalization constant in the sense of $\Gamma$-convergence. The two suitable candidates for $\sf w$ which are widely adopted read ${\sf w}(\alpha)\in\{\alpha,\alpha^2\}$, such that $c_{\sf w}\in\{\frac{8}{3},2\}$, respectively.

The two instances of formulation (\ref{RegVAF}) containing the aforementioned choices for $\sf g$ and $\sf w$ are typically termed \texttt{AT-}1 and \texttt{AT-}2 models, see Table \ref{AT}. \texttt{AT} stands for {\em Ambrosio-Tortorelli} type of regularization, see \cite{AT1992,AT1990}.

\begin{table}[ht]
\caption{Ingredients of formulation (\ref{RegVAF}).}
\centering
      \begin{tabular}{c|c|c|c}
      \hline \rule{0pt}{10pt}
        $\sf g$   &  $\sf w$ &  $c_{\sf w}$ &  Name for (\ref{RegVAF}) \vspace{0.0cm} \\
	\hline
        $(1-\alpha)^2$ &  \begin{tabular}{c} \rule{0pt}{15pt} $\alpha$ \vspace{0.1cm} \\ $\alpha^2$ \end{tabular}  
                       &  \begin{tabular}{c} \rule{0pt}{15pt} $\frac{8}{3}$ \vspace{0.1cm} \\   $2$ \end{tabular}
				       &  \begin{tabular}{c} \rule{0pt}{15pt} \texttt{AT-}1 model \vspace{0.1cm} \\ \texttt{AT-}2 model \end{tabular} \\
	\hline
      \end{tabular}
\label{AT}
\end{table}

The main difference between the two models is that \texttt{AT-}1 leads to the existence of an elastic stage before the onset of fracture, whereas using \texttt{AT-}2 the phase-field variable starts to evolve as soon as the material is loaded, see e.g.\ \cite{Pham2011,Amor2009,Marigo2016} for a more detailed explanation, as well as the discussion in Section \ref{Res1-g}.

%%%%%%%%%
\subsection{Focardi regularization}
\label{Foc_reg}
In Focardi \cite{Focardi2001}, the Ambrosio-Tortorelli regularization of the free-discontinuity problem is generalized so as to include an orientation-dependent surface energy, and the related $\Gamma$-convergence result is obtained. 
Using the notation in (\ref{VAF}) and (\ref{RegVAF}), the unregularized energy functional considered by Focardi and its regularized (phase-field) counterpart read
\begin{equation}
\mathcal{E}(\bm u,\Gamma_c): = 
\int_{\Omega\setminus\Gamma_c} 
\Psi(\bm\varepsilon(\bm u)) \, \mathrm{d}{\bf x}
+  G_0\int_{\Gamma_c} \varphi(\bm n)\, \mathrm{d}\mathcal{H}^{d-1},
\label{VAF_f}
\end{equation}
and
\begin{equation}
\mathcal{E}(\bm u,\alpha): = 
 \int_\Omega
{\sf g}(\alpha)\Psi(\bm\varepsilon(\bm u)) \, \mathrm{d}{\bf x}
+ G_0\int_\Omega 
\left(\frac{1}{\ell q}{\sf w}(\alpha)+\frac{\ell^{p-1}}{p}\varphi^p(\nabla\alpha)\right)
\mathrm{d}{\bf x},
\label{RegVAF_f}
\end{equation}
respectively. In the above, $\bm n$ is the normal unit vector to $\Gamma_c$%\footnote{More precisely, if $\Gamma_c$ is viewed as a two sided curve $\Gamma_c^\pm\subset\partial(\Omega\setminus\Gamma_c)$ it is possible to define the corresponding {\em outward} and  {\em inward} unit normal vectors $\bm n^\pm$ 
%\begin{minipage}{1.0\textwidth}
%	\begin{center}
%	\includegraphics[width=1.0\textwidth]{outward-inward.png}
%	\end{center}
%\end{minipage}
%and $\bm n$ is defined as the outward normal unit vector.}, 
, $\varphi:\mathbb{R}^d\rightarrow[0,+\infty)$ is a norm, $p\in(1,+\infty)$ and $q=p/(p-1)$; ${\sf w}(\alpha)=\alpha^p/{b_{\sf w}}$ with
 $b_{\sf w}:=\left(2\int_0^1 t^\frac{p}{q}\,\mathrm{d}t\right)^q$. The coupling function ${\sf g}$ is supposed to fulfill the standard properties listed in Section \ref{AT_reg}. Theorem 3.1 in \cite{Focardi2001} establishes the $\Gamma$-convergence result between $\mathcal{E}(\bm u,\alpha)$ and $\mathcal{E}(\bm u,\Gamma_c)$ under no specific assumptions for the norm $\varphi$.

For any choice of $\varphi$, one may say that (\ref{RegVAF_f}) provides a $p$-parametric family of regularizations to the formulation in (\ref{VAF_f}). We denote this family as the \texttt{Foc-}$p$ {\em models}.

Suppose now that $\varphi$ is a Euclidean norm. In this case, since $\bm n$ is a unit vector, we simply arrive at $\varphi(\bm n)\equiv1$ in (\ref{VAF_f}) --- this then becomes identical to (\ref{VAF}) --- and we also obtain $\varphi^p(\nabla\alpha)=|\nabla\alpha|^p$ in (\ref{RegVAF_f}). In the following, we term the $p$-parametric family of formulations in (\ref{RegVAF_f}) with $\varphi$ given by the Euclidean norm the {\em isotropic} \texttt{Foc-}$p$ {\em models} (when the reference to isotropy is clear from the context, the term isotropic is omitted). 

The isotropic \texttt{Foc-}$p$ model with $p>1$ can be viewed as an alternative to the Ambrosio-Tortorelli approximations \texttt{AT-}1 and \texttt{AT-}2, and, in particular, a generalization of the \texttt{AT-}2 case. In Table \ref{Foc} the isotropic \texttt{Foc-}2 and \texttt{Foc-}4 models are reported. One immediately notices that the \texttt{Foc-}2 formulation is identical to the \texttt{AT-}2 model, provided that the degradation function in (\ref{RegVAF_f}) is taken as ${\sf g}(\alpha)=(1-\alpha)^2$. The \texttt{Foc-}4 formulation turns out to be a more interesting case and is thoroughly investigated in Section \ref{Res1}. 
%In particular, from the explicit representation of $E_S(\alpha)$, it cannot be grasped which function will represent the so-called optimal phase-field profile for a fully developed crack\footnote{See Section \ref{Res1-profile} for details.}, and hence nothing can be said whether the corresponding transition zone is finite as in the \texttt{AT-}1 case, or infinite as in the \texttt{AT-}2 case. This will be dissected in Section \ref{Res1}. Secondly, as will also be shown in Section \ref{Res1}, the choice of the coupling function $\sf g$ is not straightforward. In particular, if we stick, e.g., with the polynomial representation for $\sf g$, the 'optimal' coupling function must be at least the eight-order polynomial! Owing to this, the governing functional $\mathcal E$ in (\ref{RegVAF_f}) becomes non-convex in $\alpha$ --- this is in contrast to, e.g., \texttt{Foc-}2 ($\equiv$ \texttt{AT-}2) case --- thus also requiring a suitable solution procedure that can cope with such non-convexity. %As already mentioned in the abstract and the introductory section, the method of our choice is the trust-region method, also see section \ref{Res1}.

\begin{table}[t]
\caption{Ingredients of formulation (\ref{RegVAF_f}) assuming $\varphi(a,b)=(a^2+b^2)^\frac{1}{2}$.}
\centering
      \begin{tabular}{c||c||c|c}
      \hline \rule{0pt}{10pt}
        $p$  & $\sf g$ &  $E_S(\alpha)$ &  Name for (\ref{RegVAF_f}) \vspace{0.1cm} \\
	\hline \rule{0pt}{20pt}
	    $2$  & $(1-\alpha)^2$ &  $\displaystyle G_0\int_\Omega 
                                    \left(\frac{1}{2\ell}\alpha^2+\frac{\ell}{2}|\nabla\alpha|^2\right)$ 
                              &  \begin{tabular}{c} {\em Isotropic} \\                                     \texttt{Foc-}2 ($\equiv$ \texttt{AT-}2) {\em model}
                                 \end{tabular}  \vspace{0.1cm} \\ 
    \hline \rule{0pt}{20pt}
        $4$  &  \begin{tabular}{c} To be \vspace{-0.1cm} \\ specified \vspace{-0.1cm} \\ (Section  \ref{Res1}) \end{tabular} &  $\displaystyle G_0\int_\Omega 
                            \left(\frac{3}{4\ell}\frac{\alpha^4}{b_{\sf w}}+\frac{\ell^3}{4}|\nabla\alpha|^4\right)$, $b_{\sf w}=2^{-\frac{4}{3}}$
			                  & \begin{tabular}{c} {\em Isotropic} \\                                     \texttt{Foc-}4 {\em model}
			                  \end{tabular}  \vspace{0.1cm} \\
	\hline
      \end{tabular}
\label{Foc}
\end{table}

However, the generality of (\ref{VAF_f}) and (\ref{RegVAF_f}) is significantly broader, since $\varphi$ does not have to be a Euclidean norm. In Section \ref{Res2}, we explain how the norm $\varphi$ in (\ref{VAF_f}) --- hence in (\ref{RegVAF_f}) --- as well as the parameter $p$ in (\ref{RegVAF_f}) can be chosen based on the anisotropy functions $\gamma_k$, $k\in\{2,4\}$ from (\ref{gammaK}), and thus used to model anisotropic fracture (we will find out that the most natural choice is $p=k$). This results in what we term the {\em anisotropic} \texttt{Foc-}$p$ {\em models}.

Table \ref{Mnem} summarizes the terminology introduced so far.
\begin{table}[ht]
\caption{Terminology.}
\centering
      \begin{tabular}{c|c||c}
      \hline \rule{0pt}{10pt}
        $\varphi$ in (\ref{VAF_f}) and (\ref{RegVAF_f})  
        &  Name for (\ref{RegVAF_f}) 
        & Aims at phase-field modeling of \vspace{0.1cm} \\
	 \hline \rule{0pt}{10pt}
	    Not specified  
	    &  \texttt{Foc-}$p$ {\em models}   
	    & ---  \vspace{0.1cm} \\ 
    \hline \rule{0pt}{10pt}
        Euclidean  
        &  {\em Isotropic} \texttt{Foc-}$p$ {\em models}  
        &  Isotropic fracture \vspace{0.1cm} \\
     \hline \rule{0pt}{10pt}
        $\gamma_p$-induced  
        &  {\em Anisotropic} \texttt{Foc-}$p$ {\em models} 
        & \begin{tabular}{c} Anisotropic fracture \\ with $p$-fold symmetric $G_c$                  \end{tabular} \vspace{0.1cm} \\ 
	\hline
      \end{tabular}
\label{Mnem}
\end{table}

%%%%%%%%%%%%
\subsection{Incremental variational problem}
We consider a quasi-static loading process in which the monotonically increasing imposed displacement $\bar{\bm u}_n$, with $n=1,2,...$ denoting the discrete pseudo-time step parameter, is prescribed on $\Gamma_{D,1}$. It is assumed that, during this process, the crack surface $\Gamma_c$ evolves incrementally and in an irreversible manner, that is, $\Gamma_{c,n}\supseteq\Gamma_{c,n-1}$.

With the phase-field energy functional $\mathcal{E}$ defined by (\ref{RegVAF}) or (\ref{RegVAF_f}), the state of the system at a given loading step $n\geq 1$ is given by
\begin{equation}
(\bm u,\alpha)=\mathrm{arg\,min} \{ \mathcal{E}(\bm v,\beta): \; \bm v \in V_{\bar{\bm u}_n}, \beta\in \mathcal{D}_{\alpha_{n-1}} \}.
\label{argmin0}
\end{equation}
Here
\begin{equation*}
V_{\bar{\bm u}_n}:=\{ \bm u\in W^{1,2}(\Omega;\mathbb{R}^d): \; 
\bm u=\bm 0 \; \mathrm{on} \; \Gamma_{D,0}, \; \bm u=\bar{\bm u}_n\; \mathrm{on} \; \Gamma_{D,1} \}
\end{equation*}
is the space of kinematically admissible displacements at step $n$, with $W^{1,2}(\Omega;\mathbb{R}^d)$ as the usual Sobolev space of functions with values in $\mathbb{R}^d$, and
\begin{equation*}
\mathcal{D}_{\alpha_{n-1}}:=\{ \alpha\in W^{1,p}(\Omega;\mathbb{R}),\,p>1: \; \alpha\geq\alpha_{n-1} \; \mathrm{in} \; \Omega \}
\end{equation*}
is the admissible space for $\alpha$ at step $n$, with $W^{1,p}(\Omega;\mathbb{R})$ as the usual Sobolev space of functions with values in $\mathbb{R}$ and $\alpha_{n-1}$ known from the previous step. The condition $\alpha\geq\alpha_{n-1}$ in $\Omega$ is used to enforce the {\em irreversibility} of the crack phase-field evolution. It is the backward difference quotient form of $\dot{\alpha}\geq0$ in $\Omega$. Also, in $\mathcal{D}_{\alpha_{n-1}}$, we take $p=2$ for the \texttt{AT-}1 and \texttt{AT-}2 models in (\ref{RegVAF}), and $p>1$ for the \texttt{Foc-}$p$ model in (\ref{RegVAF_f}).

The necessary optimality conditions for $(\bm u,\alpha)\in V_{\bm u_n}\times \mathcal{D}_{\alpha_{n-1}}$ at every loading step $n\geq1$, written in partitioned form, read:
\begin{equation}
\left\{
\begin{tabular}{l}
${\mathcal E}_{\bm u}(\bm u,\alpha;\bm v)=0 \quad \forall \; \bm v\in V_0$,   \\[0.2cm]
${\mathcal E}_\alpha(\bm u,\alpha;\beta-\alpha)\geq0 \quad \forall \; \beta\in\mathcal{D}_{\alpha_{n-1}}$,
\end{tabular}
\right.
\label{Weak0}
\end{equation}
where ${\mathcal E}_{\bm u}$ and ${\mathcal E}_\alpha$ denote the directional derivatives of ${\mathcal E}$ with respect to $\bm u$ and $\alpha$, respectively. The displacement test space in \eqref{Weak0} is
\begin{equation*}
V_0:=\{ \bm u\in W^{1,2}(\Omega;\mathbb{R}^d): \; 
\bm u=\bm 0 \; \mathrm{on} \; \Gamma_{D,0}\cup\Gamma_{D,1} \}.
\end{equation*}
Conditions (\ref{Weak0}) characterize, in general, a local minimum of $\mathcal{E}$.% (or even only a local stationary point).

%%%%%%%%%%%%
\subsection{Handling of irreversibility}
\label{Rem_Irr}
In what follows, we incorporate the irreversibility constraint $\alpha\geq\alpha_{n-1}$ via simple penalization\footnote{
A detailed discussion about various available options to handle the crack phase-field irreversibility conditions, their equivalence and related numerical comparisons can be found in our earlier publications \cite{Gerasimov2019} and \cite{LDLandTG2020}.
}, thus arriving at
\begin{equation}
{\mathcal F}(\bm u,\alpha):={\mathcal E}(\bm u,\alpha) +\frac{\hat\lambda}{2}\int_\Omega{\langle\alpha-\alpha_{n-1}\rangle}_{-}^2\,
\mathrm{d}{\bf x},
\label{RegVAF_pen}
\end{equation}
where $\mathcal{E}$ is given by (\ref{RegVAF}) or (\ref{RegVAF_f}), $\langle a \rangle_{-}:=\min(0,a)$ and $\hat\lambda\in\mathbb{R}_{+}$ is the penalty parameter. In \cite{Gerasimov2019}, an analytical procedure for the 'reasonable' choice of a lower bound for $\hat\lambda$ that guarantees a sufficiently accurate enforcement of the constraint is devised. The optimal $\hat\lambda$ is given by
\begin{equation}
\hat\lambda=
\left\{
\begin{tabular}{ll}
$\displaystyle \frac{G_0}{\ell}\frac{27}{64\,\mathtt{TOL}_\mathrm{ir}^2}$, & $\mathtt{AT}$-$1$,   \\  [0.2cm]
$\displaystyle \frac{G_0}{\ell}\left( \frac{1}{\mathtt{TOL}_\mathrm{ir}^2}-1 \right)$,  & $\mathtt{AT}$-$2$,
\end{tabular}
\right.
\label{lambda_opt}
\end{equation}
where $0<\mathtt{TOL}_\mathrm{ir}\ll1$ is a user-prescribed irreversibility tolerance threshold (in \cite{Gerasimov2019}, a practical value for $\mathtt{TOL}_\mathrm{ir}$ is suggested as $0.01$).

With $\mathcal{F}$ defined by (\ref{RegVAF_pen}), the sought solution at a given loading step $n\geq 1$ is given by
\begin{equation}
(\bm u,\alpha)=\mathrm{arg\,min} \{ \mathcal{F}(\bm v,\beta): \; \bm v \in V_{\bar{\bm u}_n}, \beta\in W^{1,p}(\Omega;\mathbb{R}) \}.
\label{argmin1}
\end{equation}
The optimality condition in this case simplifies to
\begin{equation}
\left\{
\begin{tabular}{l}
${\mathcal F}_{\bm u}(\bm u,\alpha;\bm v)=0 \quad \forall \; \bm v\in V_0$,   \\[0.2cm]
${\mathcal F}_\alpha(\bm u,\alpha;\beta)=0 \quad \forall \; \beta\in W^{1,p}(\Omega;\mathbb{R})$,
\end{tabular}
\right.
\label{Weak1}
\end{equation}
where ${\mathcal F}_{\bm u}={\mathcal E}_{\bm u}$, and
\begin{equation*}
{\mathcal F}_\alpha={\mathcal E}_\alpha
+\hat\lambda\int_\Omega{\langle\alpha-\alpha_{n-1}\rangle}_{-}\beta\; \mathrm{d}{\bf x}.
\end{equation*}
Notice that in (\ref{argmin1}) and (\ref{Weak1}) we again take $p=2$ for the \texttt{AT-}1 and \texttt{AT-}2 models, and $p>1$ for the \texttt{Foc-}$p$ model.

%Finally,
%\begin{remark}
%\label{Rem_Shear}
%In the following, for the sake of simplicity but without loss of generality, we consider   the formulation in (\ref{RegVAF_pen}) adapted to the anti-plane shear situation. In this case, $\bm u=(0,0,u_z)$ and $u_z = u:\Omega\subset\mathbb{R}^2\rightarrow\mathbb{R}$, such that $\bm\varepsilon = \nabla u: \Omega\rightarrow\mathbb{R}^2$ and $\nabla u:=(\frac{\partial }{\partial x},\frac{\partial u}{\partial y})$. As a result, the elastic energy functional $\int_\Omega{\sf g}(\alpha)\Psi(\bm\varepsilon(\bm u))$ entering $\mathcal{E}$ in (\ref{RegVAF_pen}) turns into $\frac{1}{2}\int_\Omega{\sf g}(\alpha)\mu|\nabla u|^2$ with $\mu$ as the shear modulus (Lam\'e's second parameter).
%\end{remark}

%%%%%%%%%%%%%%%%%%%%%%%%%%%%%%%%%%%%%%%%%%%%%%%%%%%%%%%%
\section{The isotropic \texttt{Foc-}$4$ model}
\label{Res1}

In Section \ref{Foc_reg} we have seen that the isotropic \texttt{Foc-}$p$ models with $p>1$ can be viewed as a family of models which includes \texttt{AT-}2 as a special case. In view of the future extension to anisotropic fracture (to be addressed in Section \ref{Res2}), and in particular to the two- and four-fold anisotropy cases, we are especially interested in the isotropic \texttt{Foc-}$2$ and \texttt{Foc-}$4$ models.  Since the former (with the standard quadratic degradation function) is identical to the \texttt{AT-}2 model, whose properties are already well known, in this section we concentrate on the isotropic \texttt{Foc-}$4$ model (which we simply denote as \texttt{Foc-}$4$ model throughout this section). The corresponding energy functional is obtained from (\ref{RegVAF_f}) by setting $\varphi(a,b):=(a^2+b^2)^\frac{1}{2}$ and $p=4$, which yields
\begin{equation}
\mathcal{E}(\bm u,\alpha) = 
\int_\Omega {\sf g}(\alpha)\Psi(\bm\varepsilon(\bm u)) \, \mathrm{d}{\bf x}
+ \frac{G_0}{4}\int_\Omega 
\left(\frac{3}{b_{\sf w}}\frac{\alpha^4}{\ell}+\ell^3|\nabla\alpha|^4\right)
\mathrm{d}{\bf x},
\label{Foc4_iso}
\end{equation}
where $b_{\sf w}=2^{-\frac{4}{3}}$.

As follows, we first carry out the analysis of a one-dimensional bar under homogeneous conditions. Through this analysis we show that the standard quadratic degradation function is not suitable for the \texttt{Foc-}4 formulation, and we then derive analytically a new suitable expression for $\sf g$. Subsequently, we study the one-dimensional localization behavior and construct the so-called optimal profile of $\alpha$ induced by the fracture energy term $E_S(\alpha)$ in (\ref{Foc4_iso}). We compare it with the optimal profiles of the \texttt{AT-}1 and \texttt{AT-}2 models. The knowledge of this profile also allows us to estimate the magnitude of the penalty parameter $\hat\lambda$ in (\ref{RegVAF_pen}) to be used for the enforcement of irreversibility in computations. Finally, the new formulation obtained combining the \texttt{Foc-}4 model and the new degradation function is tested on a simple anti-plane shear numerical setup.

%%%%%%%%%%%%
\subsection{One-dimensional homogeneous behavior}
\label{Res1-g}
The main tool for assessing the fundamental properties of a fracture phase-field formulation --- in particular, the impact of ingredients such as $\sf g$ --- is the analysis of a one-dimensional elastic bar under tension \cite{Pham2011, Marigo2016,Borden2014,Kuhn2015}. In the following, we recall the corresponding results for the \texttt{AT-}1 and \texttt{AT-}2 models, and apply the same analysis to understand the role of the degradation function and to devise a suitable one for the \texttt{Foc-}4 model.

Following \cite{Pham2011}, we consider a one-dimensional elastic bar of length $L$, clamped at one end ($u(0)=0$) and subjected to an imposed displacement $\bar{u}$ at the other end ($u(L)=\bar{u}$). The elastic energy density function reads $\Psi(\epsilon)=\frac{1}{2}E\epsilon^2$ with $E$ as the Young's modulus. In this pure tensile loading case, the phase-field irreversibility constraint is not necessary, hence we can set $\hat\lambda=0$. The strong form of the equilibrium equations in $(0,L)$ reads
\begin{equation}
\bar{\sigma}^\prime(x)=0 \quad\mbox{with}\quad \bar{\sigma}={\sf g}(\alpha)\bar{\epsilon},
\label{sig}
\end{equation}
(regardless of the model), whereas the damage criterion reads
\begin{equation}
\frac{1}{2}{\sf g}^\prime(\alpha)\frac{\bar{\epsilon}^2}{\ell}
+\frac{1}{c_{\sf w}}\left(\frac{{\sf w}^\prime(\alpha)}{\ell}
-2\ell\alpha^{\prime\prime}(x)\right)=0,
\label{evol_1}
\end{equation}
for the \texttt{AT-}1 and \texttt{AT-}2 models, and
\begin{equation}
\frac{1}{2}{\sf g}^\prime(\alpha)\frac{\bar{\epsilon}^2}{\ell}
+3\displaystyle\left(\frac{1}{b_{\sf w}}\frac{\alpha^3}{\ell}
-\ell^3[\alpha^\prime(x)]^2\alpha^{\prime\prime}(x)\right)=0,
\label{evol_2}
\end{equation}
for the \texttt{Foc-}4 model. Here, we use the re-scaled quantities $\bar{\epsilon}=\epsilon/\sqrt{\frac{G_0}{E\ell}}$ and $\bar{\sigma}=\sigma/\sqrt{\frac{G_0E}{\ell}}$.

\begin{figure}[!ht]
\begin{center}
\includegraphics[width=1.0\textwidth]{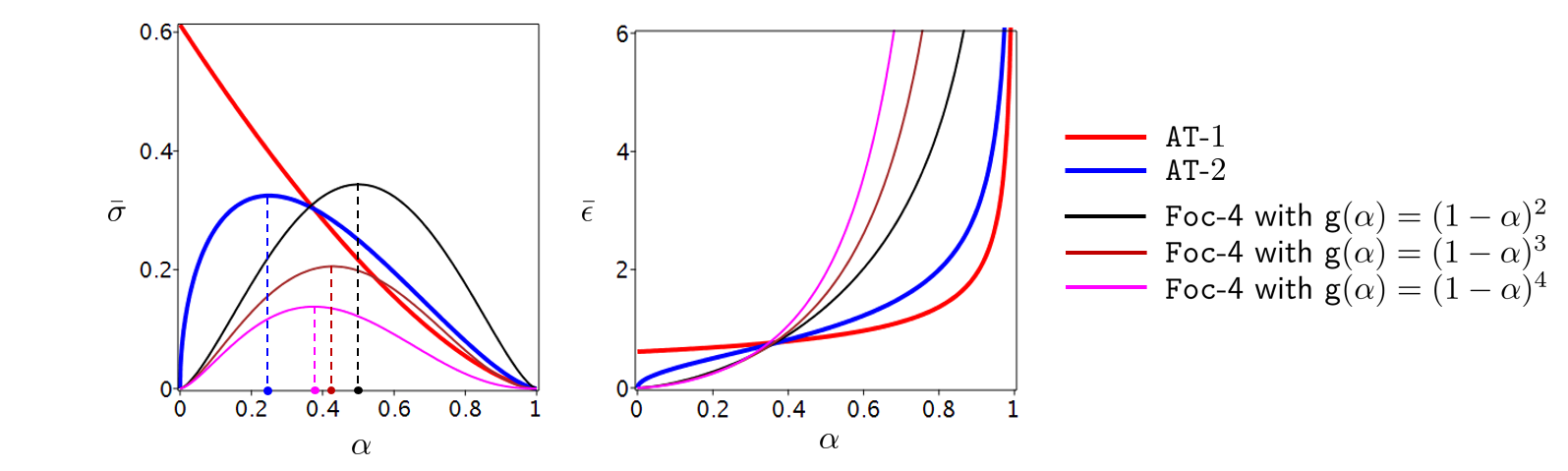}
\end{center}
\caption{Spatially homogeneous response of a one-dimensional elastic bar under tension in terms of $\bar{\epsilon}$-$\alpha$ and $\bar{\sigma}$-$\alpha$ curves, obtained using the \texttt{AT-}1 and \texttt{AT-}2 models, as well as the \texttt{Foc-}4 model with different degradation functions.}
\label{Fig:7}
\end{figure}

Let us focus on the study of the homogeneous solution, where the damage and the stress and strain fields are uniform along the bar \cite{Pham2011, Marigo2016}. In this case, due to the disappearance of the spatial derivatives of $\alpha$, equations (\ref{sig})-(\ref{evol_2}) provide the explicit $\bar{\epsilon}$-$\alpha$ and $\bar{\sigma}$-$\alpha$ relations:
\begin{itemize}
    \item for the \texttt{AT-}1 model,
    \begin{equation}
    \bar{\epsilon}=\sqrt{\frac{3}{8}\frac{1}{1-\alpha}}, \quad
    \bar{\sigma}=\sqrt{\frac{3}{8}}(1-\alpha)^\frac{3}{2},
    \label{at1}
    \end{equation}
\end{itemize}
\begin{itemize}
    \item for the \texttt{AT-}2 model,
    \begin{equation}
    \bar{\epsilon}=\sqrt{\frac{\alpha}{1-\alpha}}, \quad
    \bar{\sigma}=\alpha^\frac{1}{2}(1-\alpha)^\frac{3}{2},
    \label{at2}
    \end{equation}
\end{itemize}
\begin{itemize}
    \item for the \texttt{Foc-}4 model,
    \begin{equation}
    \bar{\epsilon}=\sqrt{-\frac{6}{b_{\sf w}}\frac{\alpha^3}{{\sf g}^\prime(\alpha)}}, \quad
    \bar{\sigma}={\sf g}(\alpha)\sqrt{-\frac{6}{b_{\sf w}}\frac{\alpha^3}{{\sf g}^\prime(\alpha)}}.
    \label{foc4}
    \end{equation}
\end{itemize}
In the first two cases, we already substituted for $\sf g$ the standard quadratic expression, whereas we are considering a generic $\sf g$ for the third case. 
Figure \ref{Fig:7} depicts the $\bar{\epsilon}$-$\alpha$ and $\bar{\sigma}$-$\alpha$ plots for the \texttt{AT-}1 and \texttt{AT-}2 models, as well as for the \texttt{Foc-}4 model with ${\sf g}(\alpha)=(1-\alpha)^t$, $t\in\{2,3,4\}$.

The curves obtained with the \texttt{AT-}1 and \texttt{AT-}2 models are well known. From them it can be inferred that the \texttt{AT-}1 model features an initial linearly-elastic phase of behavior, after which damage starts to occur. The peak value of the stress is reached for a value of $\alpha$ given by $\alpha_\mathrm{cr}=0$. Conversely, the \texttt{AT-}2 model has no purely elastic phase, as the evolution of $\alpha$ starts from the onset of loading and the peak value of the stress is reached for $\alpha_\mathrm{cr}=0.25$.

The curves corresponding to the \texttt{Foc-}4 model are qualitatively similar to those of the \texttt{AT-}2 model. However, the non-linearity prior to the peak stress is even more pronounced and the value of  $\alpha_\mathrm{cr}$ is even larger. Indeed, for ${\sf g}(\alpha)=(1-\alpha)^2$ it is $\alpha_\mathrm{cr}=0.5$. If the polynomial degree of the degradation function is increased, $\alpha_\mathrm{cr}$ is reduced, however $\alpha$ reaches $1$ at increasingly large values of $\bar{\epsilon}$. 

\begin{figure}[!ht]
\begin{center}
\includegraphics[width=1.0\textwidth]{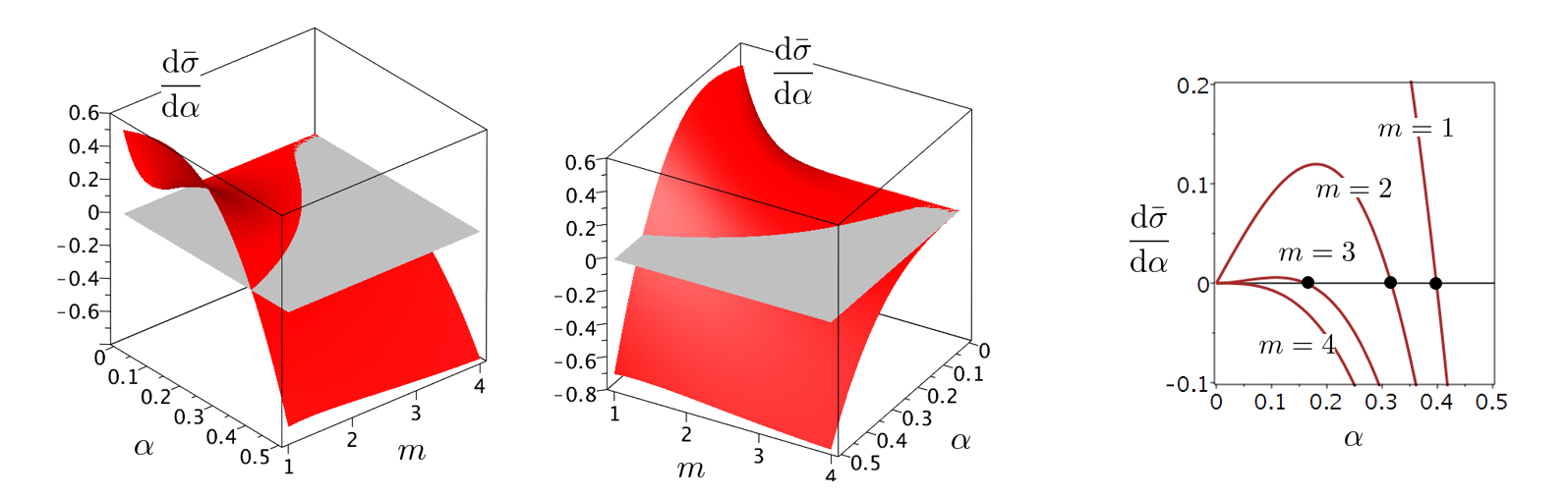}
\end{center}
\caption{The first derivative of $\bar{\sigma}$ with respect to $\alpha$ as a function of $\alpha\in[0,0.5]$ and the polynomial parameter $m\in\{1,2,3,4\}$ in (\ref{g_m}).}
\label{Fig:8}
\end{figure}

We now aim at devising a degradation function for the \texttt{Foc-}4 model  that, in addition to fulfilling the basic properties of degradation functions (see Section \ref{AT_reg}), leads to a $\bar{\epsilon}$-$\alpha$ behavior similar to the one of the \texttt{AT-}1 model. We thus enforce the following property\footnote{Clearly, the representation on the right-hand side of (\ref{cond_1}) is not unique. Even if we keep the polynomial representation at the denominator, another option that preserves the same behavior of the right-hand side is e.g. $(1-\alpha)^m$. However, it can be shown that the corresponding $m$-parametric family of $\sf g$ does not fulfil ${\sf g}^\prime(1)=0$. As a result, $\alpha$ ceases to be automatically bounded from above by 1, and incorporation of the constraint $\alpha\leq1$ in $\Omega$ within the minimization problem for $\mathcal{E}$ in (\ref{Foc4_iso}) becomes necessary.}
\begin{equation}
-\frac{\alpha^3}{{\sf g}^\prime(\alpha)}
\overset{!}{=}\frac{s}{1-\alpha^m}, \quad \alpha\in(0,1),
\label{cond_1}
\end{equation}
where $s\in\mathbb{R}$ and $m\in\mathbb{N}$, $m\geq1$. Solution of the differential equation in (\ref{cond_1}), with the boundary conditions ${\sf g}(0)=1$ and ${\sf g}(1)=0$, reads
\begin{equation}
{\sf g}(\alpha)=
1-\left(1+\frac{4}{m}\right)\alpha^4
+\frac{4}{m}\alpha^{m+4}, \quad m\geq1.
\label{g_m}
\end{equation}
Secondly, for $\bar{\sigma}$ given by (\ref{foc4}) with ${\sf g}(\alpha)$ as in (\ref{g_m}), we seek $m\geq1$ to yield $\frac{\mathrm{d}\bar{\sigma}}{\mathrm{d}\alpha}\leq0$ in $\alpha\in[0,1]$. Figure \ref{Fig:8} illustrates $\frac{\mathrm{d}\bar{\sigma}}{\mathrm{d}\alpha}$ as a function of $(\alpha,m)$. It can be clearly seen that the desired property for this quantity is fulfilled for $m=4$, yielding     
\begin{equation}
{\sf g}(\alpha)=(1-\alpha^4)^2.
\label{g_8}
\end{equation}

\begin{figure}[!ht]
\begin{center}
\includegraphics[width=1.0\textwidth]{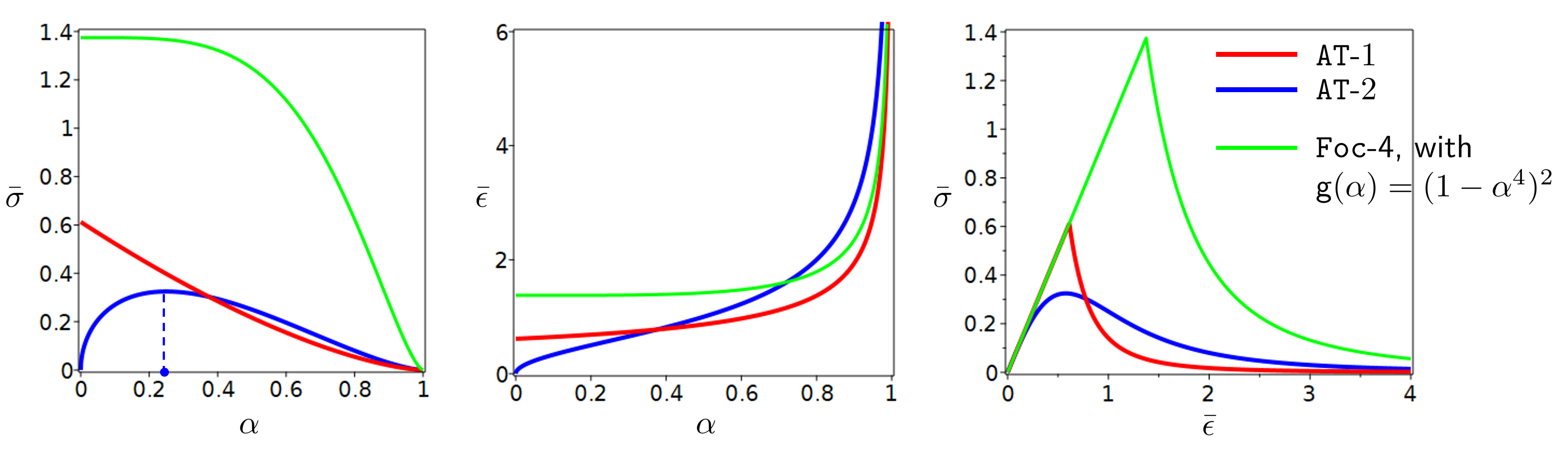}
\end{center}
\caption{Spatially homogeneous response of a one-dimensional elastic bar under tension in terms of $\bar{\epsilon}$-$\alpha$, $\bar{\sigma}$-$\alpha$ and $\bar{\sigma}$-$\bar{\epsilon}$ curves, obtained using the \texttt{AT-}1 and \texttt{AT-}2 models, as well as the \texttt{Foc-}4 model with the degradation function in (\ref{g_8}).}
\label{Fig:9}
\end{figure}

Figure \ref{Fig:9} shows the $\bar{\epsilon}$-$\alpha$, $\bar{\sigma}$-$\alpha$ and $\bar{\sigma}$-$\bar{\epsilon}$ relations for the \texttt{Foc-}4 model with the degradation function ($\ref{g_8}$), also in comparison with the \texttt{AT-}1 and \texttt{AT-}2 cases. It is clear that the constructed eighth-order polynomial function in (\ref{g_8}) leads to the desired behavior.% in terms of all $\bar{\epsilon}$-$\alpha$, $\bar{\sigma}$-$\alpha$ and $\bar{\sigma}$-$\bar{\epsilon}$ relations. 

\subsection{One-dimensional localization behavior}
\label{Res1-profile}
\begin{figure}[ht]
\begin{center}
\includegraphics[width=1.0\textwidth]{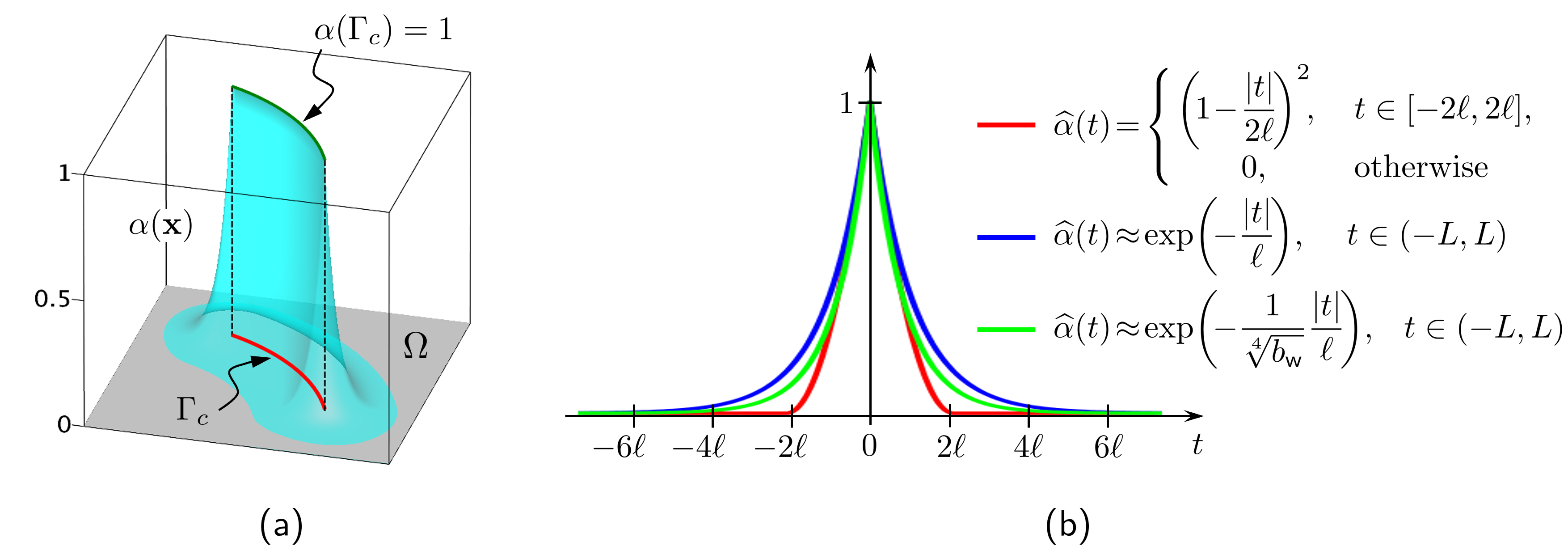}
\end{center}
\caption{Optimal phase-field profile corresponding to a fully-developed crack $\Gamma_c$ in two dimensions (a) and in one dimension for the \texttt{AT-}1, \texttt{AT-}2 and \texttt{Foc-}$4$ models (b).}
\label{Fig:5}
\end{figure}

If $\Gamma_c\subset\Omega$ is a fully developed crack, the phase-field profile $\alpha:\Omega\rightarrow[0,1]$ that represents the regularized counterpart of $\Gamma_c$ can be obtained by minimizing the fracture surface energy functional $E_S$ on a suitable admissible set, namely 
\begin{equation}
\alpha=\mathrm{arg\,min}
\left\{ E_S(\beta), \;
\beta\in W^{1,p}(\Omega;\mathbb{R}),\,p>1: \, \beta(\Gamma_c)=1, \,
\beta\geq0 \right\},
\label{Optim}
\end{equation}
see Figure \ref{Fig:5}a for an illustrative sketch in two dimensions. Using a slicing argument, one may resort to a one-dimensional setting in (\ref{Optim}): we assume $\Omega:=(-L,L)$ with $L\gg\ell$ such that $\Gamma_c$ is represented by the point $t=0$, and we also denote $\alpha:=\widehat{\alpha}(t)$. In this setting, the minimizer $\widehat{\alpha}$ can be constructed explicitly.

The strong formulation of the boundary value problem to compute $\widehat{\alpha}$ for the \texttt{AT-}1 and \texttt{AT-}2 models is given by
\begin{equation}
\left\{
\begin{tabular}{rl}
$-\ell^2\widehat{\alpha}^{\prime\prime}
+\frac{1}{2}{\sf w}^\prime(\widehat{\alpha})\geq0$ & in $(-L,L)$,\vspace{0.1cm} \\
$\widehat{\alpha}(0)=1$, & \\ 
$\widehat{\alpha}^\prime(\pm L)\geq0$. & 
\end{tabular}
\right.
\label{bvp_at12}
\end{equation}
Note that in (\ref{bvp_at12}) the inequalities hold for \texttt{AT-}1 and the corresponding equalities for \texttt{AT-}2. The solutions representing the optimal phase-field profiles read
\begin{equation}
\widehat{\alpha}(t)=
\left\{
\begin{tabular}{cc}
$\displaystyle
\left(1- \frac{|t|}{2\ell}\right)^2$, & $t\in[-2\ell,2\ell]$,   \\[0.2cm]
$0$, & otherwise,
\end{tabular}
\right.
\label{AT1_optim}
\end{equation}
for \texttt{AT-}1, and
\begin{equation}
\widehat{\alpha}(t)\approx\exp\left(-\frac{|t|}{\ell}\right), \quad t\in(-L,L),
\label{AT2_optim}
\end{equation}
for \texttt{AT-}2, see e.g. \cite{Gerasimov2019} for details. 

For the \texttt{Foc-}$4$ model, the optimal profile must be computed from
\begin{equation}
\left\{
\begin{tabular}{rl}
$-\ell^4(\widehat{\alpha}^{\prime})^2\widehat{\alpha}^{\prime\prime}
+\frac{1}{b_{\sf w}}\widehat{\alpha}^3=0$ & in $(-L,L)$,\vspace{0.1cm}   \\ 
$\widehat{\alpha}(0)=1$, & \\ 
$\widehat{\alpha}^\prime(\pm L)=0$, & 
\end{tabular}
\right.
\label{bvp_foc4}
\end{equation}
and is straightforwardly obtained as \footnote{
Assuming a general solution of the form $\widehat{\alpha}(t)=c_1\exp(-k|t|/\ell)+c_2\exp(k|t|/\ell)$, with $k$ as an unknown parameter and $c_1,c_2\in\mathbb{R}$, and substituting it into the differential equation, one obtains $k=1/\sqrt[4]{b_{\sf w}}$. Using the boundary conditions, $c_1$ and $c_2$ are derived. With the assumption $\ell\ll L$, the simplified approximate result in (\ref{Foc4_optim}) follows immediately.
}
\begin{equation}
\widehat{\alpha}(t)\approx\exp\left(-\frac{1}{\sqrt[4]{b_{\sf w}}}\frac{|t|}{\ell}\right), \quad t\in(-L,L).
\label{Foc4_optim}
\end{equation}

The plots of $\widehat{\alpha}$ in (\ref{AT1_optim}), (\ref{AT2_optim}) and (\ref{Foc4_optim}) are shown in Figure \ref{Fig:5}b.  It is evident that the \texttt{AT-}$2$ and the \texttt{Foc-}$4$ models have very similar optimal profiles. A practical outcome of this observation is that, when using the penalty method to enforce irreversibility, we can choose the penalty parameter $\hat\lambda$ for the \texttt{Foc-}$4$ formulation using the same estimate valid for the \texttt{AT-}$2$ formulation, see eq. (\ref{lambda_opt}).

%%%%%%%%%%%%
\subsection{Anti-plane shear test results}

%In this section, we consider the isotropic \texttt{Foc-}$4$ model in combination with the standard quadratic degradation function which is commonly used with the \texttt{AT-}1 and \texttt{AT-}2 models. 

%We will show that this function is not suitable for the isotropic (and, as a result, for the induced anisotropic) \texttt{Foc-}$4$ model. 

%More precisely, with ${\sf g}(\alpha)=(1-\alpha)^2$ used in, (\ref{Foc4_iso}) possesses two undesired properties. First, it is a lack of purely elastic stage, when evolution of a crack phase-field $\alpha$ starts immediately from the beginning of loading and prior to the peak load. The second one is a spatial non-zero accumulation of $\alpha$ outside the localization zone. Such a behaviour is very similar to the one of the \texttt{AT-}2 model, but in the corresponding \texttt{Foc-}4 case, both shortcomings are much more pronounced.

%\paragraph{Illustrative numerical example.} 

\begin{figure}[ht]
\begin{center}
\includegraphics[width=1.0\textwidth]{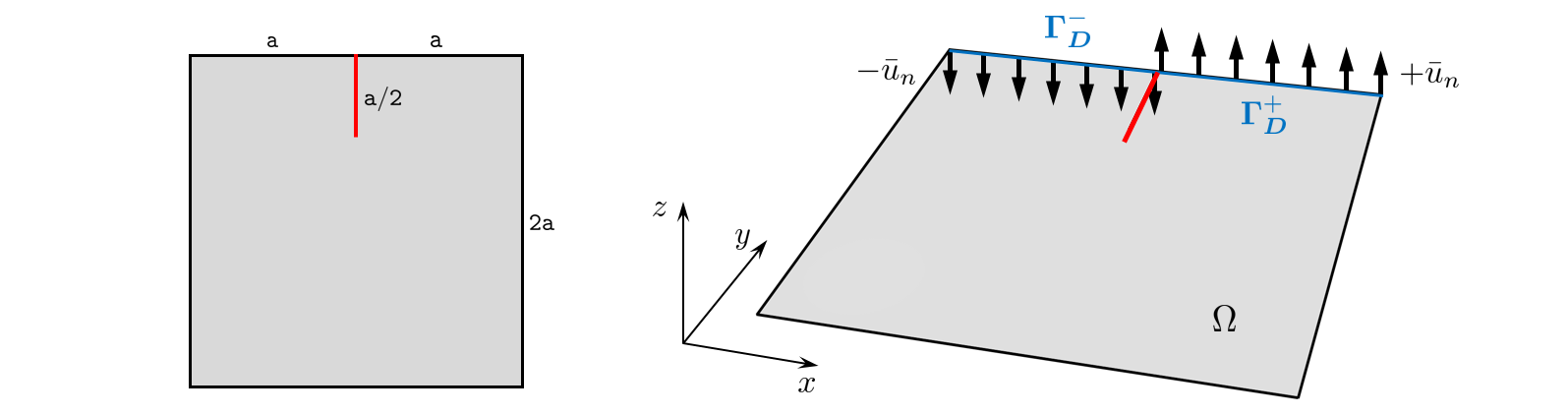}
\end{center}
\caption{Geometry and anti-plane shear loading setup.}
\label{Fig:s0}
\end{figure}

We now test the formulation consisting of the \texttt{Foc-}$4$ model with the new degradation function in (\ref{g_8}) on an anti-plane shear loading numerical setup, see Figure \ref{Fig:s0}. A two-dimensional rectangular domain $(0,\mathtt{2a})\times(0,\mathtt{2a})$ containing a slit along $\{\mathtt{a}\}\times(\mathtt{3a/2},\mathtt{2a})$ is subjected to incremental {\em anti-plane} displacements $\mp \bar{u}_n$, $n\geq1$ applied on $\Gamma_{D^-}:=(0,\mathtt{a})\times\{\mathtt{2a}\}$ and $\Gamma_{D^+}:=(\mathtt{a},\mathtt{2a})\times\{\mathtt{2a}\}$, respectively.

The loading conditions imply $\bm u=(0,0,u_z)$ with $u_z = u:\Omega\subset\mathbb{R}^2\rightarrow\mathbb{R}$, such that $\bm\varepsilon = \nabla u: \Omega\rightarrow\mathbb{R}^2$ and $\nabla u=(\frac{\partial u}{\partial x},\frac{\partial u}{\partial y})$. As a result, $\Psi(\bm\varepsilon(\bm u))=\frac{1}{2}\mu|\nabla u|^2$ with $\mu$ as the shear modulus (Lam\'e's second parameter). In the computations, we furthermore set $\mathtt{a}=1$, $\mu=1$, $\ell=\frac{2\mathtt{a}}{50}=0.04$, $G_0=1$. As already mentioned, the degradation function is taken as $\sf g(\alpha)=(1-\alpha)^2$. Also, we set $\hat\lambda=\frac{G_0}{\ell}\left( \frac{1}{\mathtt{TOL}_\mathrm{ir}^2}-1 \right)$ with $\mathtt{TOL}_\mathrm{ir}=0.01$, see Remark \ref{Rem_Irr}. The applied incremental displacement is given by $\bar{u}_n=n\Delta\bar{u}$, $n=1,...,\frac{3}{2\Delta\bar{u}}$, with $\Delta\bar{u}=0.1$ as the loading increment. We employ the numerical package FreeFem++ \cite{FreeFem}. Both the displacement field $u$ and the crack phase-field $\alpha$ are approximated using $P_1$-triangles. The finite element mesh is a pre-adapted one, which is refined in the region of $\Omega$ where crack propagation is expected. The minimum and maximum characteristic mesh sizes are given by $(h_\mathtt{min},h_\mathtt{max}):=(\frac{1}{5}\ell,\ell)$, where $h_\mathtt{min}$ and $h_\mathtt{max}$ stand for the mesh size inside and outside of the refined region, respectively. 

\begin{figure}[!ht]
\begin{center}
\includegraphics[width=1.0\textwidth]{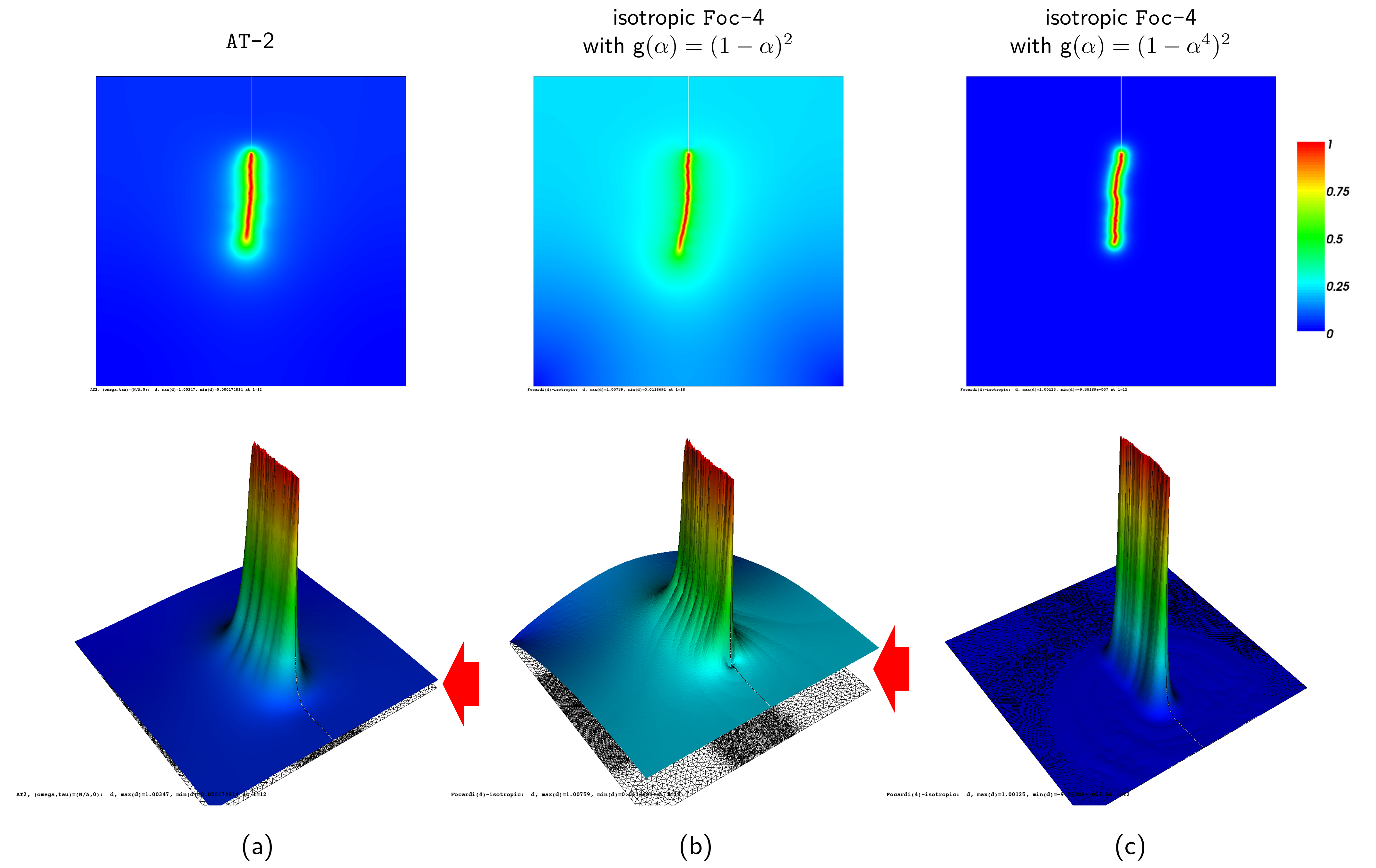}
\end{center}
\caption{Crack phase field (in two perspectives) obtained at the same post-peak loading step using the \texttt{AT-}2 model with quadratic degradation function (a), the \texttt{Foc-}4 model with quadratic degradation function (b) and the \texttt{Foc-}4 model with the degradation function in (\ref{g_8}) (c). The gap between the zero level and the non-zero $\min(\alpha)$ accumulated outside the localization zone for (a) and (b) is evidenced by an arrow.}
\label{Fig:6}
\end{figure}

Figure \ref{Fig:6}a-b visualizes the $\alpha$ fields computed with the \texttt{AT-}2 and \texttt{Foc-}4 models with quadratic degradation function at the same loading step after the peak load. It is evident that, in both cases, $\alpha$ does not vanish outside of the localization zone, which is a direct consequence of the accumulation of damage since the onset of loading discussed in Section  \ref{Res1-g}. For the given setup at the chosen loading step, it is $\min(\alpha)\approx0.05$ for \texttt{AT-}2  and $\min(\alpha)\approx0.25$ for \texttt{Foc-}4. Interestingly, both values are significantly lower than the value of $\alpha_\mathrm{cr}$ obtained for the one-dimensional homogenous responses of the respective models, see Section \ref{Res1-g}. Finally, Figure \ref{Fig:6}c shows results for the \texttt{Foc-}4 model with the degradation function in (\ref{g_8}), where no spurious damage accumulation outside of the localization region is obtained.

%Our final comment in this section regards a visible lack of symmetry in the computed crack propagation path in either case of the \texttt{AT-}2 model, as well as the \texttt{Foc-}4 model with the correct $\sf g$. We attribute this to non-convexity of the corresponding governing energy functionals\footnote{In case of the \texttt{AT-}2 model, functional $\mathcal{E}$ is separately convex with respect to $\bm u$ and $\alpha$, but it is non-convex with respect to $(\bm u,\alpha)$. For the \texttt{Foc-}4 model with the eighth-order polynomial $\sf g$ used in, already non-convexity of $\mathcal{E}$ with respect to $\alpha$ is proved, see Appendix A.2 for details.}, and the possible resulting solution non-uniqueness, see Appendix \ref{Ap1}. Similar instances of a non-symmetric fracture pattern computed for the problems with the symmetry in geometry and loading setup have been reported in the literature, see, e.g., \cite{Bourdin2000,Bourdin2007a,Bourdin2008,Amor2009,LDLandTG2020}. In our recent paper on stochastic phase-field modeling of brittle fracture \cite{TGstoch2020}, the question of characterization and description of all possible solutions occurring due to non-convex energy is addresses and studied in depth.

%%%%%%%%%%%%
%\subsection{Trust region method as a necessary solution strategy}
%Energy functional $\mathcal{E}$ in (\ref{Foc4_iso}) becomes non-convex when the correct function ${\sf g}(\alpha)=(1-\alpha^4)^2$ is incorporated. ...

%%%%%%%%%%%%%%%%%%%%%%%%%%%%%%%%%%%%%%%%%%%%%%%%%%%%%%%%
\section{The anisotropic \texttt{Foc-}$2$ and \texttt{Foc-}$4$ models}
\label{Res2}
In this section, we develop the anisotropic counterparts of the isotropic \texttt{Foc-}2 and \texttt{Foc-}4 formulations. To this end, we suitably choose the norm $\varphi$ in (\ref{VAF_f}) --- and, as a result, in (\ref{RegVAF_f}) --- based on the knowledge of the anisotropy functions $\gamma_k(\theta)$, $k\in\{2,4\}$ in (\ref{gamma2}) and (\ref{gamma4}).
In particular, $\gamma_2$ is used to induce the norm in the \texttt{Foc-}2 formulation, whereas $\gamma_4$ is employed for defining $\varphi$ in the \texttt{Foc-}4 formulation. We term the resulting formulations {\em anisotropic \texttt{Foc-}$2$ and \texttt{Foc-}$4$ models}, see Table \ref{Mnem}.

%%%%%%%%%%%%
\subsection{Derivation of the anisotropic models}
\begin{figure}[!ht]
\begin{center}
\includegraphics[width=1.0\textwidth]{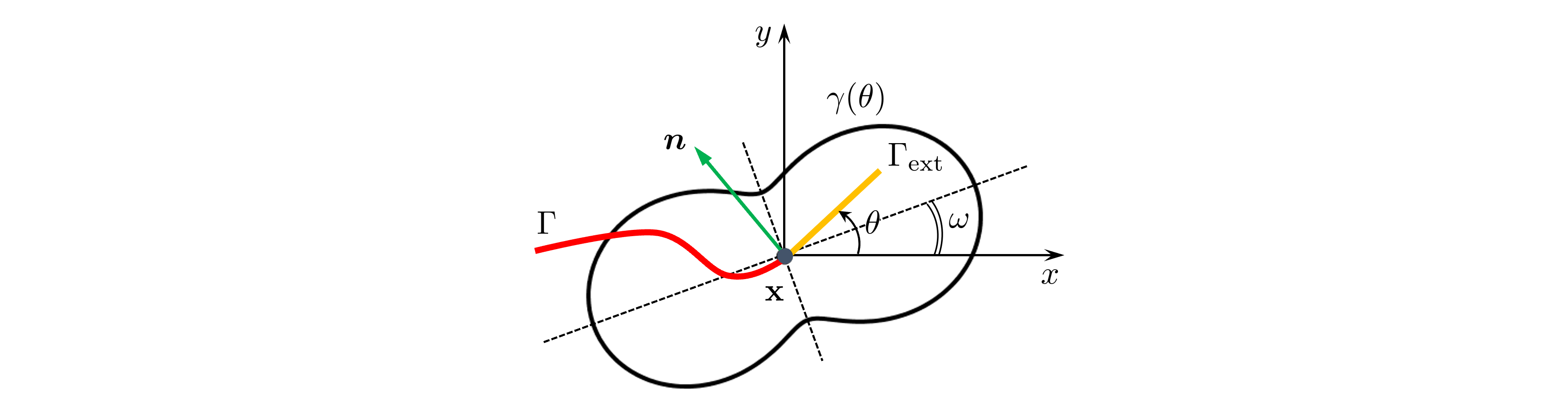}
\end{center}
\caption{Sketch relating the angular coordinate $\theta$ and the unit normal vector $\bm n$.}
\label{Fig:3}
\end{figure}

Since $\varphi$ in (\ref{VAF_f}) is assumed to be a function of the unit normal $\bm n=(n_x,n_y)$, whereas the anisotropy functions $\gamma_k$ take as argument the polar angle $\theta$, the first straightforward step is to express $\theta$ via the components of $\bm n$.
Let $\Gamma$ represent a pre-existing crack, point ${\bf x}\in\Gamma$ be the coordinate of its tip, and $\Gamma_\mathrm{ext}$ be a possible extension of $\Gamma$ starting from ${\bf x}$. Also, let us define the Cartesian and polar coordinate systems associated to $\bf x$, as well as the unit normal $\bm n$ to $\Gamma_\mathrm{ext}$ at $\bf x$. The angle $\theta$ is defined as the angle between the positive direction of the $x$-axis and the crack extension $\Gamma_\mathrm{ext}$, see Figure \ref{Fig:3}. With this definition, $\theta$ is related to the components of $\bm n$ by $n_x=\cos(\theta+\frac{\pi}{2})=-\sin(\theta)$ and $n_y=\sin(\theta+\frac{\pi}{2})=\cos(\theta)$. 
Substituting in (\ref{gamma2}) and (\ref{gamma4}), we obtain
\begin{equation}
\gamma_2(\bm n)=
    n_x^2+n_y^2+\tau
    \left[ 
    \cos(2\omega)(n_y^2-n_x^2)-2\sin(2\omega)n_x n_y
    \right],
\label{gamma2-n}
\end{equation}
and
\begin{equation}
\gamma_4(\bm n)=
    (n_x^2+n_y^2)^2+\tau
    \left[
    \cos(4\omega)(n_y^4-6n_x^2n_y^2+n_x^4)
    -4\sin(4\omega)n_x n_y(n_y^2-n_x^2)
    \right].
\label{gamma4-n}
\end{equation}
where we also exploited the following representations of the unity

\begin{equation}
1:=\left\{
\begin{tabular}{cc}
$n_x^2+n_y^2$ & in $\gamma_2$,   \\[0.1cm]
$(n_x^2+n_y^2)^2$ & in $\gamma_4$,
\end{tabular}
\right.
\label{unity}
\end{equation}
We can now introduce the mappings $\varphi:\mathbb{R}^2\rightarrow[0,+\infty)$ for two- and four-fold anisotropy, which respectively read:
\begin{equation}
\varphi(a,b):=
\left\{
    a^2+b^2+\tau
    \left[ 
    \cos(2\omega)(b^2-a^2)-2\sin(2\omega)ab
    \right]
\right\}^\frac{1}{2},
\label{phi2}
\end{equation}
and
\begin{equation}
\varphi(a,b):=
\left\{
    (a^2+b^2)^2+\tau
    \left[
    \cos(4\omega)(b^4-6a^2b^2+a^4)
    -4\sin(4\omega)ab(b^2-a^2)
    \right]
\right\}^\frac{1}{4}.
\label{phi4}
\end{equation}
We call $\varphi$ in (\ref{phi2}) and (\ref{phi4}) $\gamma_2$- and $\gamma_4$-induced norms, respectively, to be used in the Focardi discrete variational formulation in (\ref{VAF_f}). Noting now that the above norms $\varphi$ to be used in (\ref{VAF_f}) have the form $\varphi=f^\frac{1}{k}$, $k\in\{2,4\}$, and that the consequent terms $\varphi^p$, $p>1$ to appear in (\ref{RegVAF_f}) read $\varphi^p=f^\frac{p}{k}$, the most natural choice is to set $p\overset{!}{=}k\in\{2,4\}$. This yields
\begin{equation}
\label{phi2_buona}
\varphi^2(\nabla\alpha)=|\nabla\alpha|^2
+\tau
\left[
\cos(2\omega)(\alpha_{,y}^2-\alpha_{,x}^2)-2\sin(2\omega)\alpha_{,x}\alpha_{,y}
\right],
\end{equation}
as the ingredient of (\ref{RegVAF_f}) when $p=2$ and $\varphi$ is the $\gamma_2$-induced norm, and
\begin{equation}
\label{phi4_buona}
\varphi^4(\nabla\alpha)=|\nabla\alpha|^4
+\tau
\left[
\cos(4\omega)(\alpha_{,y}^4-6\alpha_{,x}^2\alpha_{,y}^2+\alpha_{,x}^4)
            -4\sin(4\omega)\alpha_{,x}\alpha_{,y}(\alpha_{,y}^2-\alpha_{,x}^2)
\right],
\end{equation}
as the ingredient of (\ref{RegVAF_f}) when $p=4$ and $\varphi$ is the $\gamma_4$-induced norm. We denoted $(\alpha_{,x},\alpha_{,y})$ as the components of $\nabla \alpha$.

The norm in (\ref{phi2_buona}) can be also written as

\begin{equation}
\label{phi2_tensor}
\varphi^{2}\left(\nabla\alpha\right)=\mathbf{B}\nabla\alpha\cdot\nabla\alpha=B_{ij}\alpha_{,i}\alpha_{,j}
\end{equation}
with $i,j=1,2$. Here

\begin{equation}
\label{B2}
\mathbf{B}=\mathbf{I}-\tau\mathbf{D}\qquad\qquad\mathrm{with}\qquad\mathbf{D}=\left[\begin{array}{cc}
\cos\left(2\omega\right) & \sin\left(2\omega\right)\\
\sin\left(2\omega\right) & -\cos\left(2\omega\right)
\end{array}\right]
\end{equation}
 is a symmetric positive definite second-order tensor, and $\mathbf{I}$
is the two-dimensional second-order unit tensor. 
The norm in (\ref{phi4_buona}) can be written as

\begin{equation}
\label{phi4_tensor}
\varphi^{4}\left(\nabla\alpha\right)=\mathbf{\mathbb{B}}\nabla\alpha\otimes\nabla\alpha\cdot\nabla\alpha\otimes\nabla\alpha=\mathbb{B}_{ijkl}\alpha_{,i}\alpha_{,j}\alpha_{,k}\alpha_{,l}
\end{equation}
with $i,j,k,l=1,2$. Here

\begin{equation}
\label{B4}
\mathbf{\mathbb{B}}=\mathbf{I}\otimes\mathbf{I}-\tau\mathbf{\mathbb{D}}
\end{equation}
is a positive definite fourth-order tensor exhibiting minor and major
symmetries, with components

\begin{eqnarray}
\mathbb{D}_{1111}=\mathbb{D}_{2222} & = & -\cos\left(4\omega\right)\\
\mathbb{D}_{1122}=\mathbb{D}_{2211}=\mathbb{D}_{1212}=\mathbb{D}_{2121}=\mathbb{D}_{1221}=\mathbb{D}_{2112} & = & \cos\left(4\omega\right)\\
\mathbb{D}_{1112}=\mathbb{D}_{1121}=\mathbb{D}_{1211}=\mathbb{D}_{2111} & = & -\sin\left(4\omega\right)\\
\mathbb{D}_{2221}=\mathbb{D}_{2212}=\mathbb{D}_{2122}=\mathbb{D}_{1222} & = & \sin\left(4\omega\right)
\end{eqnarray}
%Using Voigt notation, we can also transform $\mathbb{B}$ into the
%symmetric positive definite second-order tensor
%
%\begin{equation}
%\mathbf{\hat{B}=I}_{3}-\tau\mathbf{\hat{D}}\qquad\mathrm{with}\qquad\hat{\mathbf{D}}=\left[\begin{array}{ccc}
%-\cos\left(4\omega\right) & \cos\left(4\omega\right) & -\sin\left(4\omega\right)\\
%\cos\left(4\omega\right) & -\cos\left(4\omega\right) & \sin\left(4\omega\right)\\
%-\sin\left(4\omega\right) & \sin\left(4\omega\right) & \cos\left(4\omega\right)
%\end{array}\right]
%\end{equation}
%where $\mathbf{I}_{3}$ is the three-dimensional second-order unit
%tensor.

\begin{remark}
The representation of  $1$ in terms of the components $(n_x,n_y)$ of the unit vector is not unique. However, the choice given by (\ref{unity}) ---  and resulting in the appearance of the terms $a^2+b^2$ and $(a^2+b^2)^2$ in (\ref{phi2}) and (\ref{phi4}), respectively --- is the only appropriate one in this case, as it preserves an identical dimension for all terms entering the corresponding $\varphi$.
\end{remark}

\begin{remark}
The choice of the exponents $\frac{1}{2}$ and $\frac{1}{4}$ in (\ref{phi2}) and (\ref{phi4}) is a natural one, since the limiting case of $\tau=0$, with $(a,b)\neq\bm0$, turns the corresponding $\varphi$ into the Euclidean norm.
\end{remark}

\begin{remark}
The mappings in $\varphi$ in (\ref{phi2}) and (\ref{phi4}) indeed define two norms, since owing to 
\begin{equation}
\left\{
\begin{tabular}{l}
$\inf_{\omega\in[0,\pi)}(\cos(2\omega)(b^2-a^2)-2\sin(2\omega)ab)=-(a^2+b^2)$,   \\[0.1cm]
$\sup_{\omega\in[0,\pi)}(\cos(2\omega)(b^2-a^2)-2\sin(2\omega)ab)=(a^2+b^2)$,
\end{tabular}
\right.
\label{infsup_2}
\end{equation}
and
\begin{equation}
\left\{
\begin{tabular}{l}
$\inf_{\omega\in[0,\frac{\pi}{2})}(\cos(4\omega)(b^4-6a^2b^2+a^4)-4\sin(4\omega)ab(b^2-a^2))=-(a^2+b^2)^2$,   \\[0.1cm]
$\sup_{\omega\in[0,\frac{\pi}{2})}(\cos(4\omega)(b^4-6a^2b^2+a^4)-4\sin(4\omega)ab(b^2-a^2))=(a^2+b^2)^2$,
\end{tabular}
\right.
\label{infsup_4}
\end{equation}
the following norm-equivalence result holds:
\begin{equation*}
(1-\tau)|(a,b)|\leq\varphi(a,b)\leq(1+\tau)|(a,b)|,
\end{equation*}
for all $\tau\in[0,1)$ and $\omega$ belonging to the appropriate range.
\end{remark}

We can now finalize the proposed phase-field models for anisotropic fracture with two- and four-fold symmetric fracture toughness. For two-fold symmetry, i.e. for $G_c(\theta)=G_0\gamma_2(\theta)$ $=G_0\left(1+\tau\cos(2(\theta-\omega))\right)$, we propose the following energy functional:
\begin{equation}
\mathcal{E}(\bm u,\alpha):=    
\int_\Omega
(1-\alpha)^2\Psi(\bm\varepsilon(\bm u)) \, \mathrm{d}{\bf x}
+\frac{G_0}{2}\int_\Omega 
\left\{ 
\frac{\alpha^2}{\ell}
+\ell\varphi^{2}\left(\nabla\alpha\right)
\right\}  \mathrm{d}{\bf x},
\label{Foc2_aniso}
\end{equation}
with $\varphi^{2}\left(\nabla\alpha\right)$ given by (\ref{phi2_buona}) or equivalently by (\ref{phi2_tensor}), 
whereas for four-fold symmetry, i.e. for $G_c(\theta)=G_0\gamma_4(\theta)$ $=G_0\left[1+\tau\cos(4(\theta-\omega))\right]$, we propose
\begin{equation}
\mathcal{E}(\bm u,\alpha):=    
\int_\Omega
 (1-\alpha^4)^2\Psi(\bm\varepsilon(\bm u)) \,\mathrm{d}{\bf x}
+\frac{G_0}{4}\int_\Omega 
\left\{ 
\frac{3}{b_{\sf w}}\frac{\alpha^4}{\ell}
+\ell^3\varphi^{4}\left(\nabla\alpha\right)
\right\} \mathrm{d}{\bf x}.
\label{Foc4_aniso}
\end{equation}
with $\varphi^{4}\left(\nabla\alpha\right)$ given by (\ref{phi4_buona}) or equivalently by (\ref{phi4_tensor}), and where also $b_{\sf w}=2^{-\frac{4}{3}}$.
We term the formulations (\ref{Foc2_aniso}) and (\ref{Foc4_aniso}) the {\em anisotropic} \texttt{Foc-}$2$ and  \texttt{Foc-}$4$ {\em models}, respectively.

The questions of existence and uniqueness of the solution to the minimization problem for $\mathcal{E}$ in (\ref{Foc2_aniso}) and (\ref{Foc4_aniso}) is a delicate task. Indeed, it is very well known that for the isotropic case, $\min_{(\bm u,\alpha)}\mathcal{E}$ may admit infinitely many solutions, or none. In the light of the so-called alternate minimization scheme, see e.g. \cite{Bourdin2000,Bourdin2007a,Bourdin2007b,Bourdin2008}, applied to resolve $\min_{(\bm u,\alpha)}\mathcal{E}$ numerically, one typically restricts the discussion to studying the well-posedness of the problems $\min_{\bm u}\mathcal{E}$ when $\alpha$ is fixed, and $\min_{\alpha}\mathcal{E}$ when $\bm u$ is fixed. The former case is simple: the problem is well-posed due to the convexity of $\mathcal{E}$ with respect to $\bm u$. Well-posedness of the latter minimization problem with $\mathcal{E}$ given by (\ref{Foc2_aniso}) and (\ref{Foc4_aniso}) is discussed in Appendices A.1 and A.2, respectively. In the appendices, this is formulated as an $\tau$-parametric result.

\subsection{Relation to existing phase-field models for anisotropic fracture}

It is straightforward to recognize that the surface energy integral in (\ref{Foc2_aniso}) is a specific instance of the class of integrals proposed in the literature for two-fold anisotropy \cite{Hakim2005, Hakim2009, Li2019}

\begin{equation}
E_{S}\left(\alpha\right)=\frac{G_{0}}{c_{w}}\int_{\Omega}\left(\frac{w\left(\alpha\right)}{\ell}+\ell\mathbf{B}\nabla\alpha\cdot\nabla\alpha\right)
\end{equation}
where $w(\alpha)$ and $c_w$ are specialized to the \texttt{AT-}2 model. It is more interesting to compare the surface energy integral in (\ref{Foc4_aniso}) with the integral proposed in \cite{Li2019} for four-fold anisotropy (a simplified version of the model in \cite{Li2015}):

\begin{equation}
\label{Li_Maurini}
E_{S}\left(\alpha\right)=\frac{G_{0}}{c_{w}}\int_{\Omega}\left(\frac{w\left(\alpha\right)}{\ell}+\ell^{3}\mathbf{\mathbb{B}}\nabla^{2}\alpha\cdot\nabla^{2}\alpha\right)
\end{equation}
Here the fourth-order tensor $\mathbb{B}$ is a positive definite fourth-order tensor exhibiting minor and major
symmetries, just like the $\mathbb{B}$ tensor in (\ref{phi4_tensor}). However, our new formulation has the advantage of involving only first-order derivatives of $\alpha$, so that a standard $C^0$ finite element approximation can be adopted, whereas the second-order derivatives of $\alpha$ contained in (\ref{Li_Maurini}) call for $C^1$-continuous finite element approximations or mixed approaches.

\subsection{Numerical solution}
Let the energy functional $\mathcal{F}$ be as in (\ref{RegVAF_pen}) with $\mathcal{E}$ represented by (\ref{Foc4_iso}), (\ref{Foc2_aniso}) and (\ref{Foc4_aniso}). The staggered iterative process for solving $\mathcal{F}(\bm u,\alpha)\rightarrow\mathrm{min}$ at any fixed loading step $n\geq0$ consists in the following three steps: with the known initial guess $(\bm u_0,\alpha_0)$, for any staggered iteration $k\geq1$, 

\begin{enumerate}
    \item given $\bm u_{k-1}$, solve $\mathcal{F}(\bm u_{k-1},\alpha)\rightarrow\mathrm{min}$ ($\Leftrightarrow$ $\mathcal{F}_\alpha(\bm u_{k-1},\alpha;\beta)=0$) for $\alpha$, set $\alpha=:\alpha_k$,
    \item given $\alpha_k$, solve $\mathcal{F}(\bm u,\alpha_k)\rightarrow\mathrm{min}$ ($\Leftrightarrow$ $\mathcal{F}_{\bm u}(\bm u,\alpha_k;\bm v)=0$) for $\bm u$, set $\bm u=:\bm u_k$,
    \item for the computed pair $(\bm u_k,\alpha_k)$ check
    \begin{equation*}
        \mathrm{ResSTAG}_k:=\mathcal{F}_{\bm u}(\bm u_k,\alpha_k;\bm v)+\mathcal{F}_\alpha(\bm u_k,\alpha_k;\beta)\leq\mathrm{TolSTAG}.
    \end{equation*}
\end{enumerate}

As shown in Appendix A.2, both the isotropic and anisotropic \texttt{Foc-}$4$ formulations in (\ref{Foc4_iso}) and (\ref{Foc4_aniso}), respectively, are non-convex with respect to the crack phase-field variable $\alpha$. The implication is that solving $\mathcal{F}_\alpha(\bm u_{k-1},\alpha;\beta)=0$ with standard Newton-Raphson iterative procedures may lead to convergence difficulties. In the context of phase-field modeling, a sophisticated line search algorithm  was developed in \cite{Gerasimov2016}. In this paper, instead of applying the same algorithm for the case at hand, we investigate the use of the trust-region (TR) method \cite{TRmethods}, first adopted in \cite{Krause} in combination with phase-field simulations.

In the version we use, the TR method relies on the second-order Taylor expansion of $\mathcal{F}$ around a given fixed $\alpha$: 
\begin{equation*}
\mathcal{F}(\bm u,\alpha+z)
\approx
\mathcal{F}(\bm u,\alpha)
+\mathcal{F}_\alpha(\bm u,\alpha;z)
+\frac{1}{2}\mathcal{F}_{\alpha\alpha}(\bm u,\alpha;z;z)
=:M(z).
\end{equation*}
with $\mathcal{F}_{\alpha\alpha}$ as the second-order directional derivative of $\mathcal{F}$ with respect to $\alpha$. The iterative procedure of obtaining $\alpha_k$ is as follows: given $\alpha^{(i)}$, $i\geq0$ (as the initial guess, one may take $\alpha^{(0)}:=\alpha_{k-1}$), the minimization problem to be solved at every $i$ reads
\begin{equation}
M_i(z):=
\frac{1}{2}\mathcal{F}_{\alpha\alpha}(\bm u_{k-1},\alpha^{(i)};z;z)
+\mathcal{F}_\alpha(\bm u_{k-1},\alpha^{(i)};z)
+\mathcal{F}(\bm u_{k-1},\alpha^{(i)})
\rightarrow\mathrm{min},
\label{Mi}
\end{equation}
for all $z\in H^1(\Omega)$ such that $||z||_{L^m(\Omega)}\leq R_i$, $m\geq1$. Here, $R_i>0$ is the TR radius, considered as an iteration-dependent quantity. Notice also that recent $m$ must not be confused with $m$ from section \ref{Res1-g}. For the computed $z$, we obtain the {\em trial step} solution $\alpha^{(i+1)}=\alpha^{(i)}+z$ and the corresponding TR ratio
\begin{equation}
\rho_i:=\frac
{\mathcal{F}(\bm u_{k-1},\alpha^{(i)})
-\mathcal{F}(\bm u_{k-1},\alpha^{(i)}+z)}
{\mathcal{F}(\bm u_{k-1},\alpha^{(i)})
-M_i(z)}.
\label{rhoi}
\end{equation}
If $0<\eta_1<\rho_i<\eta_2<1$, the solution $\alpha^{(i+1)}$ is considered acceptable and the TR radius $R_i$ does not need to be modified. If $\rho_i\leq\eta_1$ (it can happen to be even negative), the solution is considered unacceptable, $R_i$ is reduced by factor $\gamma_1<1$ and $z$ has to be recomputed. Finally, when $\rho_i\geq\eta_2$, $\alpha^{(i+1)}$ is acceptable and $R_i$ can be enlarged by factor $\gamma_2$. Typical values for the parameters are $\eta_1=\frac{1}{4}$, $\eta_1=\frac{3}{4}$ and $\gamma_1=\frac{1}{2}$, $\gamma_2\in\{2,3\}$. The iterative solution process in $i$ is terminated once a certain criterion is achieved, and we set the last $\alpha^{(i)}$ to be $\alpha_k$. 

Further relevant details are highlighted as follows:
\begin{itemize}
\item The minimization in (\ref{Mi}) is carried out under the point-wise constraint $-R_i\leq z\leq R_i$ in $\Omega$. This is the strongest constraint of the type $||z||_{L^m(\Omega)}\leq R_i$, $m\geq1$, but is also the simplest one in terms of implementation. We incorporate it via penalization by minimizing the modified functional:
\begin{equation}
\widetilde{M}_i(z):=M_i(z)
+\frac{\lambda}{2}\int_\Omega
\left(
{\langle R_i+z\rangle}_{-}^2
+{\langle R_i-z\rangle}_{-}^2
\right)
\,\mathrm{d}{\bf x}
\rightarrow\mathrm{min},
\quad \forall\; z\in H^1(\Omega),
\label{MiPen}
\end{equation}
with $\lambda\gg 1$. In our numerical experiments, we typically take $R_0=0.01$ (this proves reasonable since the crack phase-field variable is bounded $\alpha\in[0,1]$, whereas $z$ can be viewed as a 'correction' and hence cannot be 'large'). Also, in all following examples we set the penalty parameter $\lambda$ in (\ref{MiPen}) as $10^4$.

\item To compute $z=\mathrm{arg\,min}\{\widetilde{M}_i(z): z\in H^1(\Omega)\}$ at every fixed $i\geq0$, the optimality condition $(\widetilde{M}_i)_z(z;Y)=0$ for all $Y\in H^1(\Omega)$ is used. Here, $(\widetilde{M}_i)_z$ is the first derivative of $\widetilde{M}_i$ in the direction of $Y$. Due to the Macaulay brackets in the penalty term, the problem is non-linear in $z$ and can be solved e.g. with the Newton-Raphson method. One important remark here is that the initial guess $z_0$ for the Newton-Raphson update $z_0+\Delta z$ should not be 0 in order not to solve the original badly-behaved problem $\mathcal{F}_\alpha(\bm u_{k-1},\Delta z; Y)=0$.  

\item In (\ref{rhoi}), $M_i$ is substituted by $\widetilde{M}_i$.

\item Our termination criterion for the acceptable trial step solution $\alpha^{(i+1)}$ reads as follows. First, let us notice that checking whether the residual $\mathcal{F}_\alpha(\bm u_{k-1},\alpha^{(i+1)};\beta)$ is sufficiently small is an inappropriate criterion due to non-convexity of $\mathcal{F}$ in $\alpha$: even if $\mathcal{F}(\bm u_{k-1},\alpha^{(i+1)})<\mathcal{F}(\bm u_{k-1},\alpha^{(i)})$ holds true, as required, it can very well happen -- and we indeed observed this in the 
computations -- that $\mathcal{F}_\alpha(\bm u_{k-1},\alpha^{(i+1)};\beta)\gg\mathcal{F}_\alpha(\bm u_{k-1},\alpha^{(i)};\beta)$. As a result, we opt for monitoring and checking the relative error in $\mathcal{F}$ and stop iterating in $i$ when the condition $\frac{|\mathcal{F}(\bm u_{k-1},\alpha^{(i+1)})-\mathcal{F}(\bm u_{k-1},\alpha^{(i)})|}{\mathcal{F}(\bm u_{k-1},\alpha^{(i)})}\leq \mathrm{TolPF}$ is fulfilled. In our experiments, we use  $\mathrm{TolPF}=10^{-4}$. 
\end{itemize}

%%%%%%%%%%%%
\section{Numerical experiments}
\label{NumTest}
As follows, we illustrate some numerical experiments on the presented anisotropic \texttt{Foc-}$2$ and \texttt{Foc-}$4$ formulations. Given that the former belongs to an already known family of models for two-fold anisotropy, we mainly concentrate on the latter, which for the first time describes four-fold anisotropy with a second-order model. 

We restrict ourselves to an anti-plane shear setup. In addition to being computationally less expensive due to a scalar unknown displacement field, this setup offers the advantage that the anisotropy of the elastic stiffness properties does not play a role, at least as long as we consider anisotropy classes where normal and shear stresses remain uncoupled (e.g. orthorombic or cubic).

\subsection{Numerical experiments for the anisotropic \texttt{Foc-}$2$ model}
\label{Res2-anis2}
In this section we present a numerical example for the anisotropic \texttt{Foc-}2 model in (\ref{Foc2_aniso}), with the geometry and loading setup in Figure \ref{Fig:s0}. We set $\mathtt{a}=1$, $\mu=1$, $\ell=\frac{2\mathtt{a}}{50}=0.04$. The anisotropic fracture toughness is defined by $G_0=1$, $\omega=\frac{\pi}{4}$, and $\tau=\{0.2,0.5,0.8\}$. The penalty parameter to enforce irreversibility  is set to $\hat\lambda=\frac{G_0}{\ell}\left( \frac{1}{\mathtt{TOL}_\mathrm{ir}^2}-1 \right)$ with $\mathtt{TOL}_\mathrm{ir}=0.01$. The applied incremental displacement is given by $\bar{u}_n=n\Delta\bar{u}$ with $\Delta\bar{u}=0.1$ and $n=1,...,\frac{3}{2\Delta\bar{u}}$. Both $u$ and $\alpha$ are approximated using $P_1$-triangles. The finite element mesh is identical for all computations, which only differ by the value of $\tau$. The mesh is pre-refined in the region where crack propagation is expected and the characteristic mesh sizes in the refined and in the coarser regions are $(h_\mathtt{min},h_\mathtt{max}):=(\frac{1}{5}\ell,\ell)$.

\begin{figure}[!ht]
\begin{center}
\includegraphics[width=1.0\textwidth]{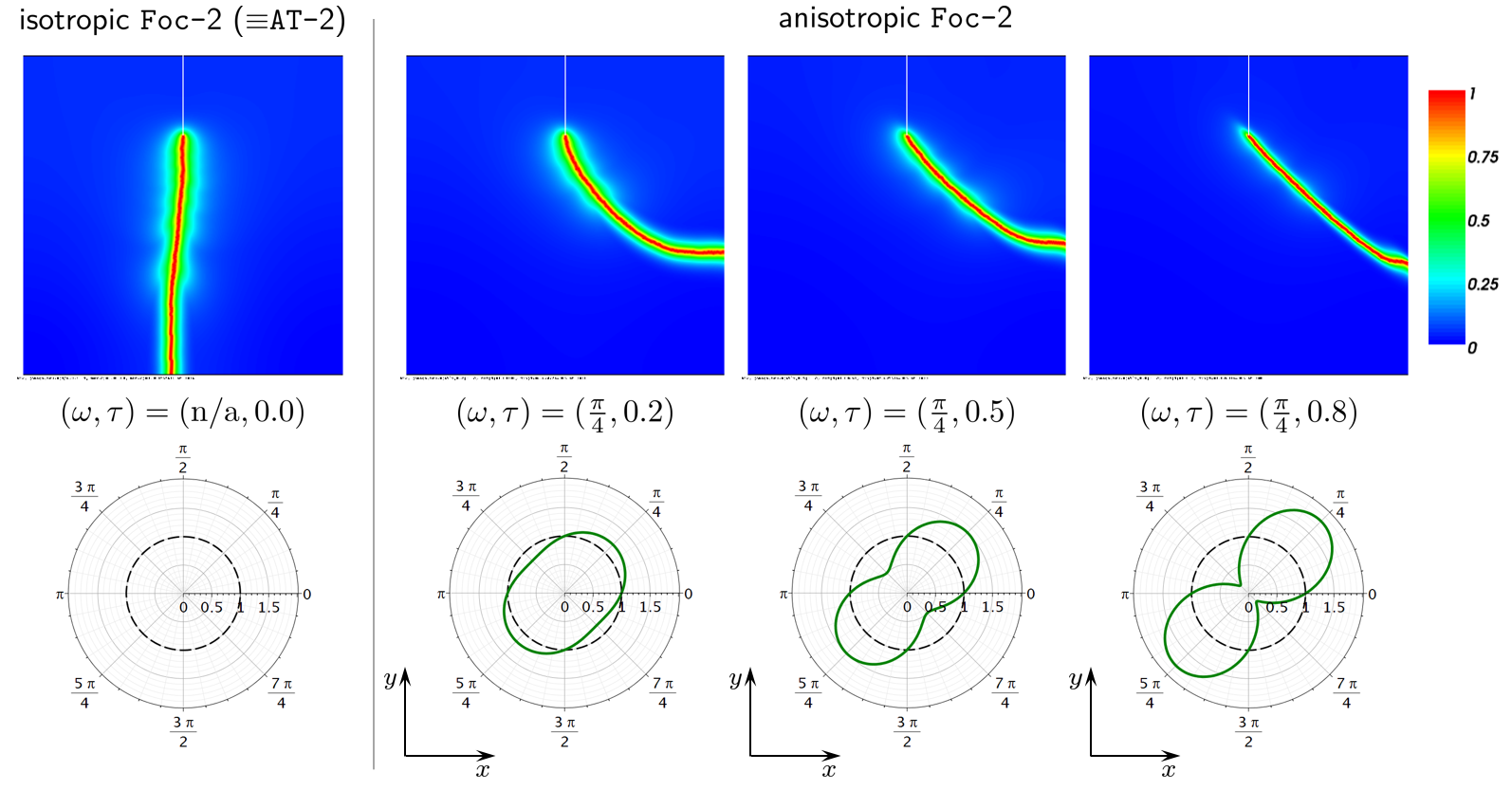}
\end{center}
\caption{Phase-field contour plots at the final loading step for the setup in Figure \ref{Fig:s0}. The anisotropy function is $\gamma_2(\theta)$ with the specified values of $(\omega,\tau)$, as depicted in the lower inserts.}
\label{Fig:s0_Foc2}
\end{figure}

The results are presented in Figure \ref{Fig:s0_Foc2}, where for comparison we also show results obtained with the isotropic \texttt{Foc-}2 ($\equiv$\texttt{AT-}2) formulation. It can be observed that, for a non-zero anisotropy strength $\tau$, the crack path deviates from the vertical direction obtained in the isotropic case. As expected, for all considered values of $\tau$, the crack tends to propagate along the weakest material direction (in this case, at the angle $-\frac{\pi}{4}$), and the increase of $\tau$ makes this effect more pronounced: already for $\tau=0.5$ (such that $\frac{1+\tau}{1-\tau}=3$) the angle of propagation is very close to $-\frac{\pi}{4}$ at the initial loading stage and then changes due to boundary effects, whereas for the largest $\tau=0.8$ (yielding $\frac{1+\tau}{1-\tau}=9$) the propagation angle aligned with the weak direction $-\frac{\pi}{4}$ persists until complete failure.

%%%%%%%%%%%%
\subsection{Numerical experiments for the anisotropic \texttt{Foc-}$4$ model}
\label{Res2-anis4}
\begin{figure}[!ht]
\begin{center}
\includegraphics[width=1.0\textwidth]{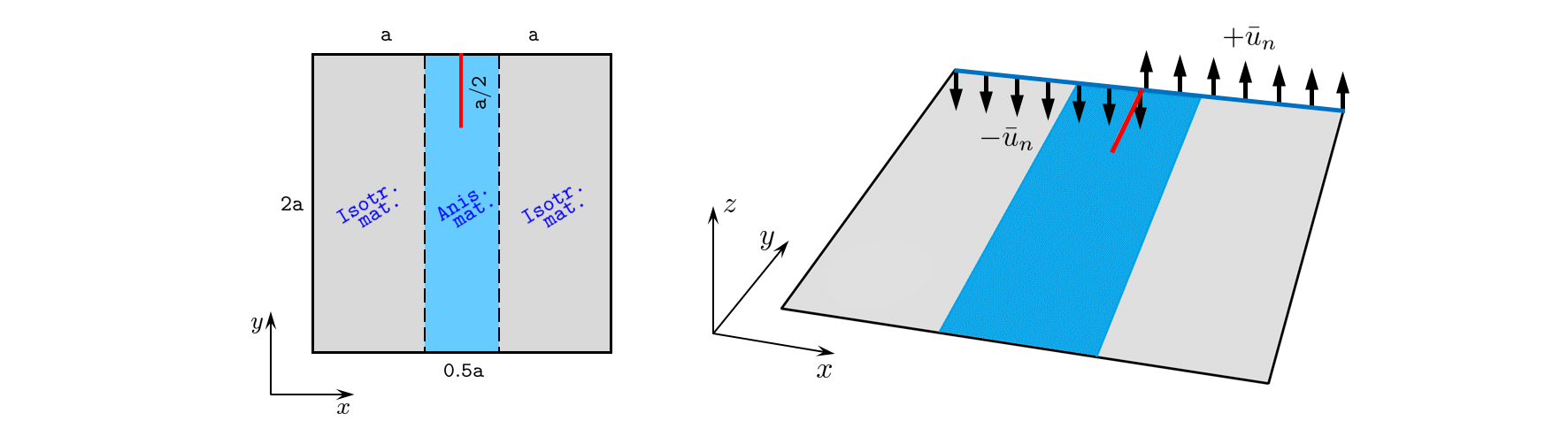}
\end{center}
\caption{Example 1: geometry and loading setup.}
\label{Fig:ex1}
\end{figure}

\begin{figure}[!ht]
\begin{center}
\includegraphics[width=1.0\textwidth]{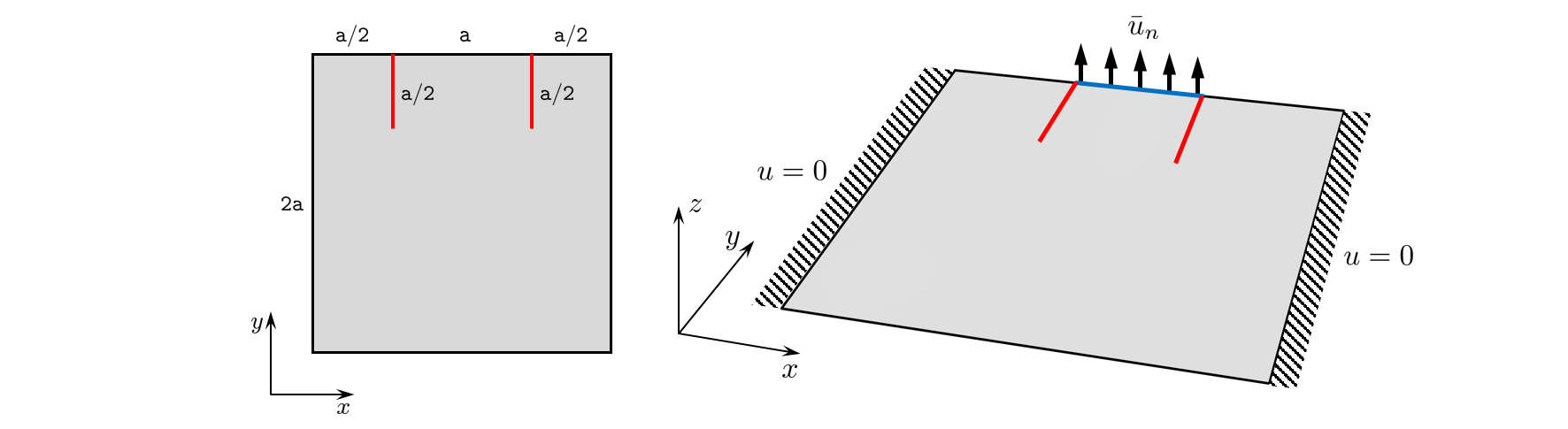}
\end{center}
\caption{Example 2: geometry and loading setup.}
\label{Fig:ex2}
\end{figure}

\begin{figure}[!ht]
\begin{center}
\includegraphics[width=1.0\textwidth]{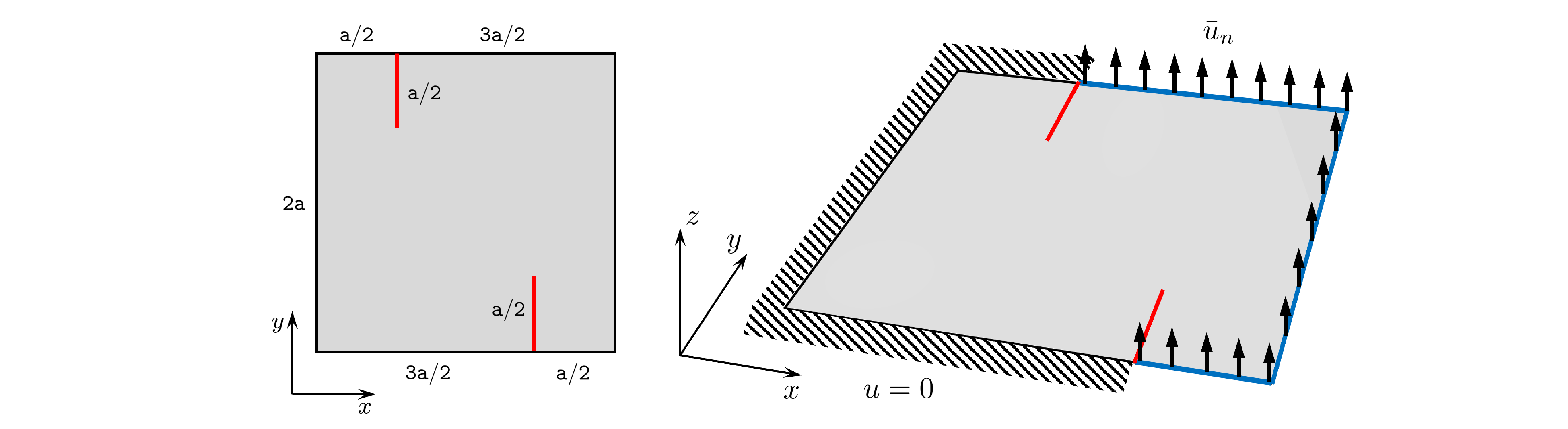}
\end{center}
\caption{Example 3: geometry and loading setup.}
\label{Fig:ex3}
\end{figure}

In this section we perform numerical tests for the anisotropic \texttt{Foc-}$4$ formulation in (\ref{Foc4_aniso}), using the anti-plane shear loading setups in Figures \ref{Fig:ex1} (example 1),  \ref{Fig:ex2} (example 2) and \ref{Fig:ex3} (example 3). 
In terms of geometry, all the examples are represented by a square $\mathtt{2a}\times\mathtt{2a}$ domain containing one slit (example 1), or two slits (examples 2 and 3). The slit length is $\mathtt{a/2}$ in all cases. The applied incremental displacement  is given by $\bar{u}_n=n\Delta\bar{u}$, $\Delta\bar{u}=0.1$, where $n=1,...,20$. All examples share the following geometry, material and model parameters: $\mathtt{a}=1$, $\ell=\frac{2\mathtt{a}}{50}=0.04$, $\mu=1$, $G_0=1$ and $\hat\lambda=\frac{G_0}{\ell}\left( \frac{1}{\mathtt{TOL}_\mathrm{ir}^2}-1 \right)$ with $\mathtt{TOL}_\mathrm{ir}=0.01$.

%%%%%% example 1, 2 and 3 -- setup

In example 1, the domain consists of a central strip of material featuring four-fold anisotropic fracture toughness, and two side strips of isotropic material with $G_c(\theta)\equiv100G_0$. For simplicity, the shear modulus $\mu$ is identical for both materials. For this example, the strongest material direction $\omega$ and the anisotropy strength parameter $\tau$ are fixed to $0$ and  $0.5$, respectively. In examples 2 and 3, the domain consists entirely of anisotropic material, and for the fixed parameter $\tau=0.5$ we compare two values of $\omega\in\{0,\frac{\pi}{4}\}$.

The displacement field $u$ and the crack phase-field $\alpha$ are approximated by $P_1$-triangles. We pre-refine our mesh in the region where crack propagation is expected. In example 1, we compare results with $(h_\mathtt{min},h_\mathtt{max}):=(\frac{1}{4}\ell,\ell), (\frac{1}{5}\ell,\ell), (\frac{1}{6}\ell,\ell)$. In examples 2 and 3, we only consider $(h_\mathtt{min},h_\mathtt{max}):=(\frac{1}{5}\ell,\ell)$.

%We are now in a position to analyse and comment on our numerical findings in each example. Before doing that, let us notice that in all figures where the computed crack phase-field pattern is presented, we also always depict the anisotropy function $\gamma_4(\theta)$ along with the specified values of $(\omega,\tau)$. This aims at easing identification of the corresponding results. 
 
%%%%%% example 1 -- results

\subsubsection{Example 1}
In this example, the idea of the bi-material arrangement with strong direction $\omega=0$ is to induce a crack evolution along the weakest material directions within the middle strip, accompanied by crack reflections at the boundaries with the two side strips (where the crack cannot propagate due to the significantly larger fracture toughness). 

\begin{figure}[!ht]
\begin{center}
\includegraphics[width=1.0\textwidth]{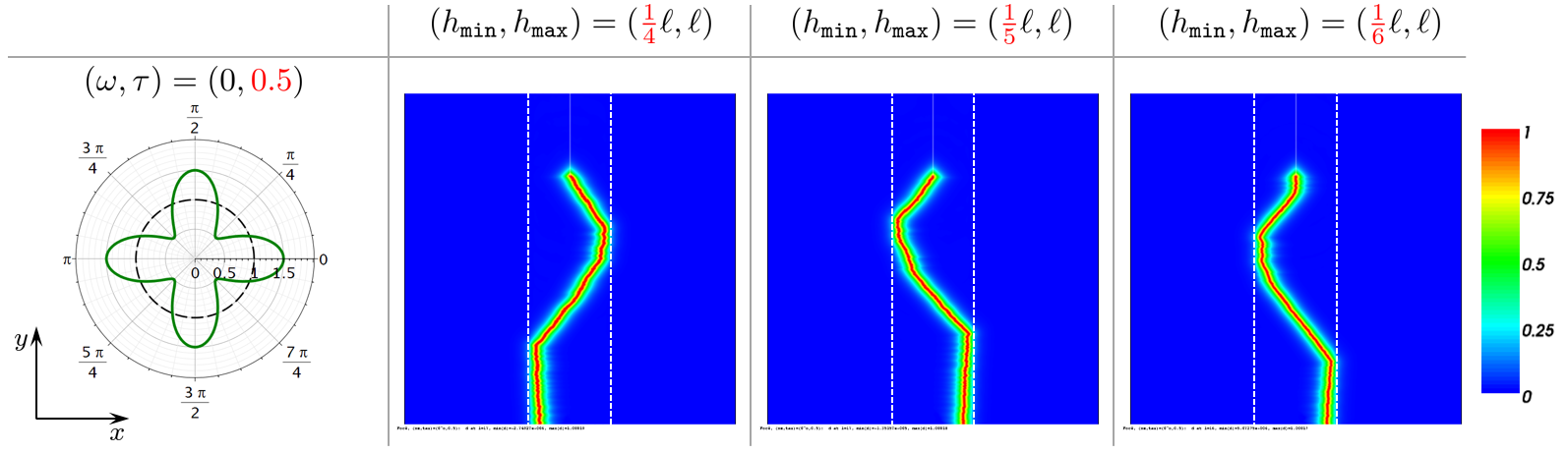}
\end{center}
\caption{Example 1: Phase-field contour plots at the final loading step. For the material in the central strip, the anisotropy function is $\gamma_4(\theta)$ with the specified values of $(\omega,\tau)$, as depicted in the left insert. Results for three different meshes are compared.}
\label{Fig:ex1_res1}
\end{figure}

The results are presented in Figure \ref{Fig:ex1_res1}, where the crack phase-field is computed on meshes with different element sizes. The first observation is that the actual crack propagation pattern only partially corresponds to the expectations. Let us start by looking at results for $(h_\mathtt{min},h_\mathtt{max}):=(\frac{1}{4}\ell,\ell)$. The crack does start propagating along the weak material direction $-\frac{\pi}{4}$. However, upon hitting the interface with the side strip of very tough material, it is not immediately deviated to the symmetric weak material direction $\frac{\pi}{4}$, but rather proceeds for a certain length along the bimaterial interface and then deviates to $\frac{\pi}{4}$. When the crack hits the other bimaterial interface, once again no immediate reflection takes place, and the crack proceeds along the bimaterial interface until reaching the end of the specimen. 

\begin{figure}[!ht]
\begin{center}
\includegraphics[width=1.0\textwidth]{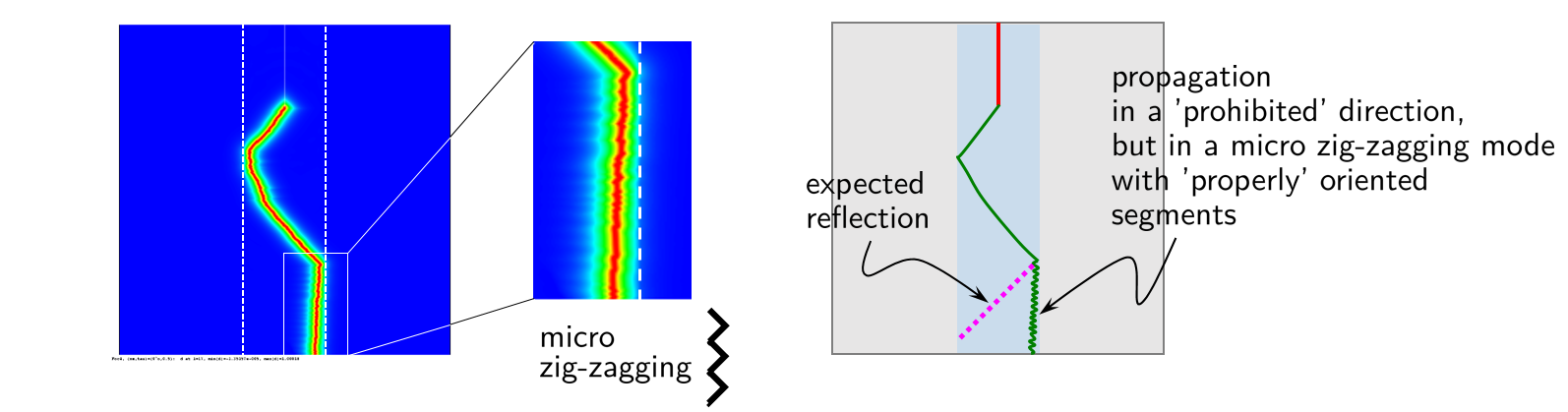}
\end{center}
\caption{Example 1: closer examination of results from Figure \ref{Fig:ex1_res1} for $(h_\mathtt{min},h_\mathtt{max}):=(\frac{1}{5}\ell,\ell)$. We illustrate how the crack does not reflect from the interface along one of the weakest material directions, as expected, but propagates along the strongest material direction in a micro zig-zagging mode.}
\label{Fig:ex1_res2a}
\end{figure}

A second observation stems from the comparison between the results obtained with the three meshes. It appears that the crack patterns are qualitatively similar. Also, it is evident that the results obtained for the last two meshes are roughly symmetric with respect to those of the first mesh. Due to symmetry, for each crack evolution two symmetric patterns are fully equivalent, and which one is obtained numerically is determined by the mesh, as well as by parameters related to the solution algorithm (including e.g. loading increments, thresholds and tolerances), and by round-off errors. Note that multiple solutions are possible already in isotropic fracture, see \cite{TGstoch2020}, but the range of possibilities appears to be even wider in the anisotropic case.

A third observation is the following. As mentioned above, we expected the crack to be reflected immediately upon hitting the bimaterial interface, and not to follow it as seen in Figure \ref{Fig:ex1_res1}. Indeed, the bimaterial interface is aligned with the strongest material direction. Interestingly, even though the crack macroscopically propagates  in this 'prohibited' direction, at a closer look it is seen to consist of many small segments following the two weak material directions. We refer to this phenomenon as 'microscopic (or micro) zig-zagging' and evidence it in Figure \ref{Fig:ex1_res2a}. Here the term 'microscopic' is intended to refer to a size scale of the order of the size of a finite element, as opposed to the macroscopic scale which relates to the characteristic size of the domain under investigation. 

\begin{figure}[!ht]
\begin{center}
\includegraphics[width=1.0\textwidth]{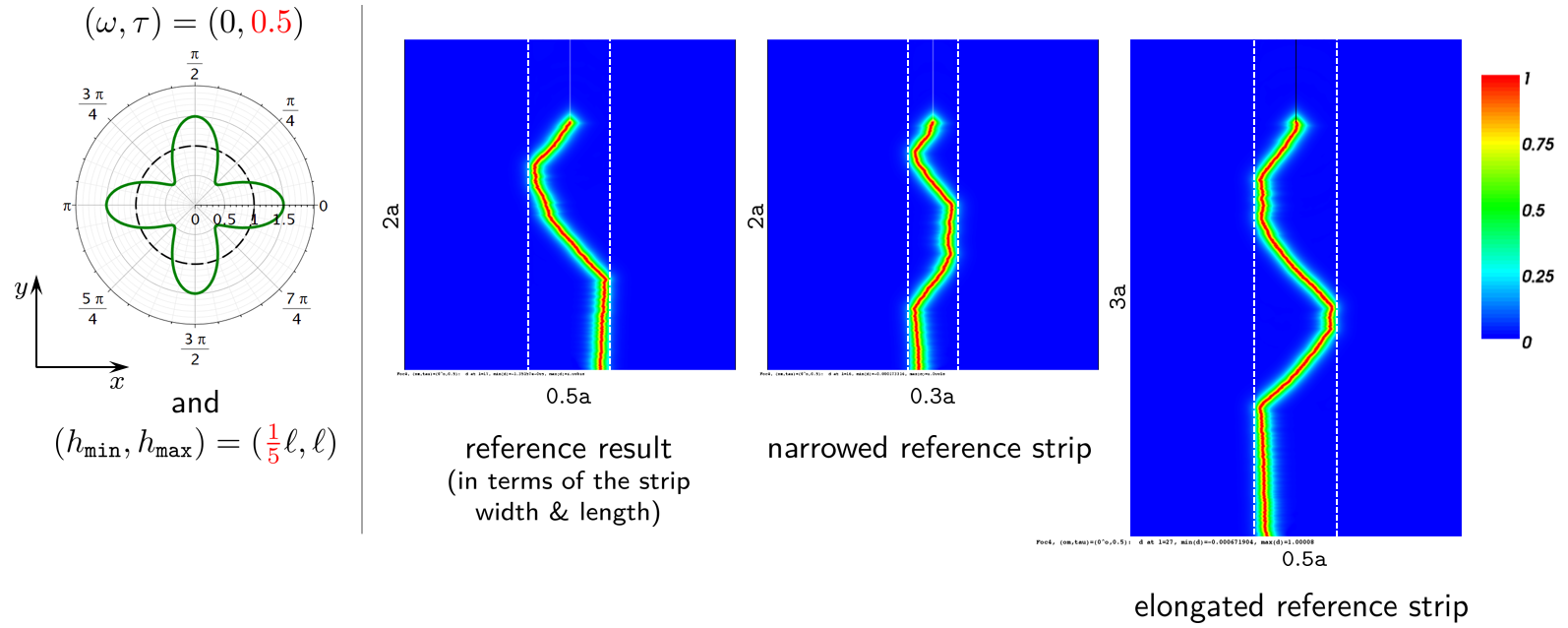}
\end{center}
\caption{Example 1: Results for a narrower (middle plot) or longer (right plot) middle strip with respect to the specimen in Figure \ref{Fig:ex1_res1}.}
\label{Fig:ex1_res3}
\end{figure}

In order to observe whether the number of reflections at the bimaterial interface is influenced by the length-to-width ratio of the middle strip, we numerically test two additional specimens where the middle strip is either narrower or longer than in the specimen of Figure \ref{Fig:ex1_res1}, see Figure \ref{Fig:ex1_res3}. Indeed, when the length-to-width ratio of the strip is increased by either decreasing its width or increasing its length, the number of reflections increases by one. Also in these patterns, the crack partially follows the bimaterial interface between subsequent reflections and along the final part of the propagation, which locally induces once again a micro zig-zagging phenomenon.

%\begin{figure}[!ht]
%\begin{center}
%\includegraphics[width=1.0\textwidth]{ex1_res2_b.png}
%\end{center}
%\caption{Example 1: result from Figure \ref{Fig:ex1_res1} for the mesh size $(h_\mathtt{min},h_\mathtt{max}):=(\frac{1}{6}\ell,\ell)$; illustration of how crack propagates after initiation not in the weakest, but in a 'prohibited' direction in a micro zig-zagging mode.}
%\label{Fig:ex1_res2b}
%\end{figure}

%%%%%% example 2 -- results
\subsubsection{Example 2}
\begin{figure}[!ht]
\begin{center}
\includegraphics[width=1.0\textwidth]{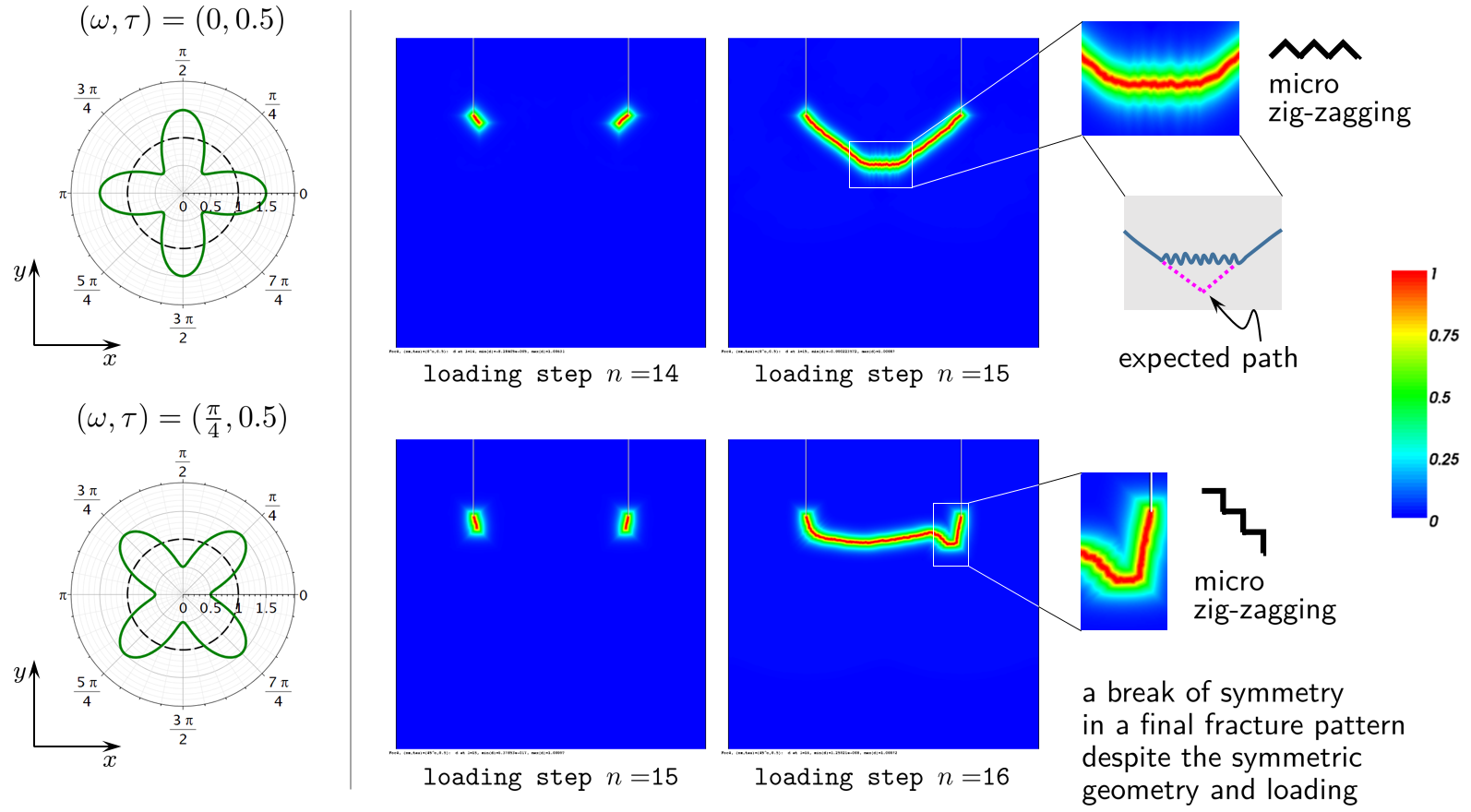}
\end{center}
\caption{Example 2: Phase-field contour plots at the final loading step. The anisotropy function is $\gamma_4(\theta)$ with the specified values of $(\omega,\tau)$, as depicted in the left inserts.}
\label{Fig:ex2_res}
\end{figure}

In this example, due to the presence of two initial cracks, we expect both cracks to propagate and to join in the middle of the specimen, following a pattern depending on the choice of the strongest material directions. Results are shown in Figure \ref{Fig:ex2_res} for $\omega=0$ and $\omega=\frac{\pi}{4}$. In the first case, the simplest expected crack pattern would consist in the propagation of both initial cracks along $+\frac{\pi}{4}$ (for the left crack) and $-\frac{\pi}{4}$ (for the right crack), until both cracks join on the middle axis of the specimen.  The obtained results are similar to this expectation, with the exception of a final segment followed by both cracks in the horizontal (i.e. the strongest) material direction, and again featuring a micro zig-zag pattern. In the second case, the two cracks propagate following first the vertical and then approximately the horizontal direction, i.e. both weakest directions. However, also in this case some local deviations are observed, which break the symmetry of the results and introduce again micro zig-zagging patterns. These local deviations and the loss of symmetry intuitively suggest that many alternative crack patterns could be obtained through small perturbations of the system, see \cite{TGstoch2020}. 

%%%%%% example 3 -- results
\subsubsection{Example 3}
\begin{figure}[!ht]
\begin{center}
\includegraphics[width=1.0\textwidth]{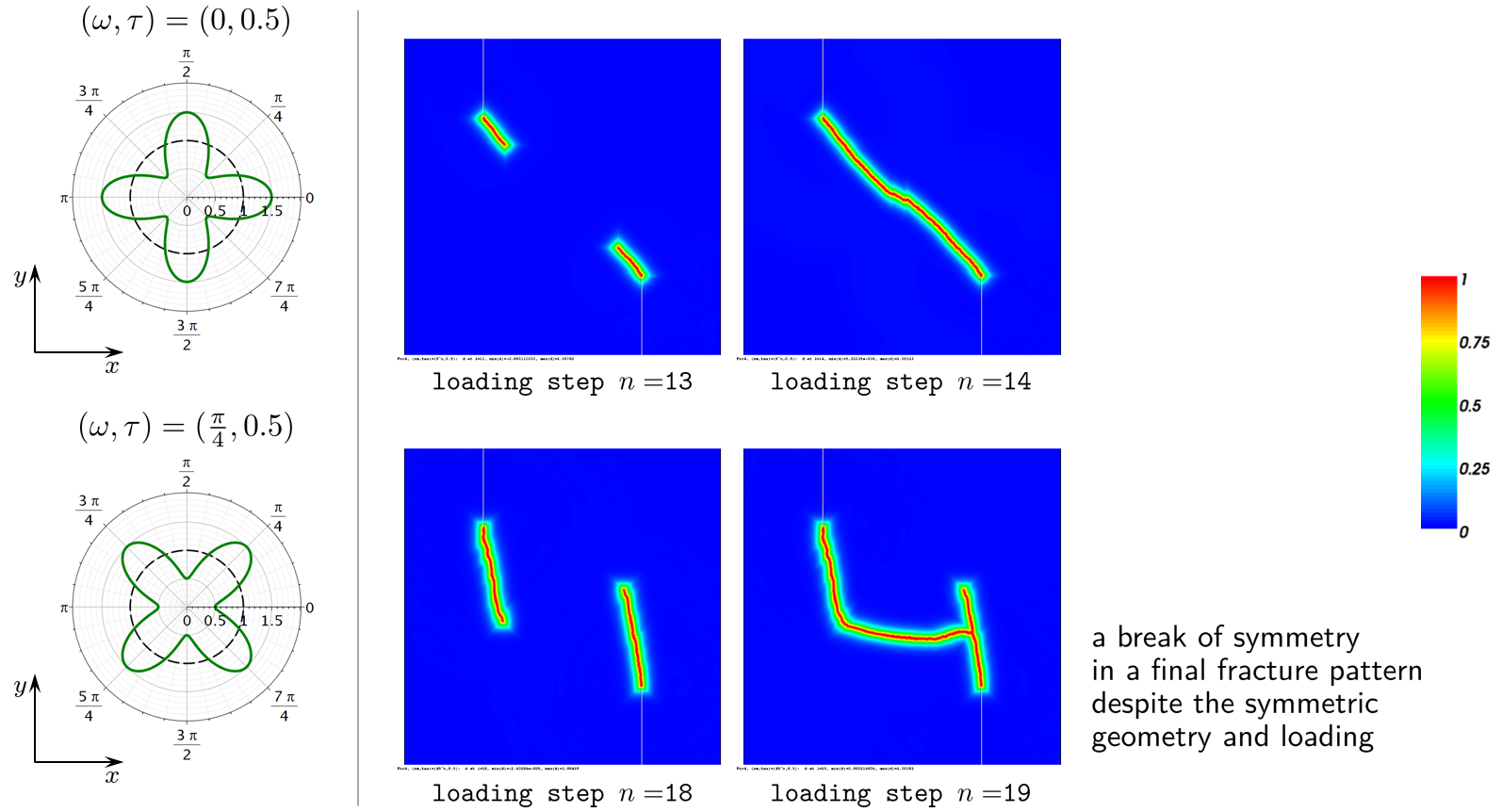}
\end{center}
\caption{Example 3: Phase-field contour plots at the indicated loading steps. The anisotropy function is $\gamma_4(\theta)$ with the specified values of $(\omega,\tau)$, as depicted in the left inserts.}
\label{Fig:ex3_A_res}
\end{figure}

Like example 2, this example features two initial cracks, which however are situated on opposite sides of the specimen. Similarly as in the previous two examples, we expect the two cracks to propagate along the weakest material directions and to merge at some point in the middle of the specimen. Once again, we examine both $\omega=0$ and $\omega=\frac{\pi}{4}$. Results are illustrated in Figure \ref{Fig:ex3_A_res}.

For $\omega=0$, the two initial cracks propagate approximately along the weakest material direction $-\frac{\pi}{4}$ and then merge in the middle of the specimen, according to the expectations. For $\omega=\frac{\pi}{4}$, this quite natural cracking pattern can no longer take place since the weakest material directions are now the vertical and the horizontal one. As a result, the two cracks propagate very closely to the vertical direction, and then merge along a direction close to horizontal. In both cases, no portion of the crack is aligned with the strongest material directions, hence no micro zig-zagging is observed. Once again the loss of symmetry in the final crack pattern for $\omega=\frac{\pi}{4}$ suggests that alternative patterns are potentially possible upon slight perturbations \cite{TGstoch2020}. 

%\begin{figure}[!ht]
%\begin{center}
%\includegraphics[width=1.0\textwidth]{ex3(B)_res.png}
%\end{center}
%\caption{Example 3 with the loading setup (B): Phase-field contour plots at the indicated loading steps. The anisotropy function is $\gamma_4(\theta)$ with the specified values of $(\omega,\tau)$, as depicted in the left inserts.}
%\label{Fig:ex3(B)_res}
%\end{figure}

%%%%%%%%%%%%%%%%%%%%%%%%%%%%%%%%%%%%%%%%%%%%%%%%%%%%%%%%
\section{Conclusions}
\label{Conclusions}

Based on the family of $\Gamma$-convergent regularizations proposed by Focardi \cite{Focardi2001}, we developed two variational phase-field models for fracture in anisotropic materials featuring two-fold and four-fold symmetric fracture toughness. Since both proposed models are of second order, as opposed to the previously available fourth-order models for four-fold symmetric fracture toughness, they do not require higher-continuity basis functions nor mixed variational principles for finite element discretization. However, the model for four-fold symmetry features a fracture energy functional that is non-convex with respect to the phase-field variable and thus requires a numerical solution algorithm able to cope robustly with non-convex minimization. In our numerical examples, we adopt the trust region method and assess the performance of the new models through several numerical examples simulating anisotropic fracture under anti-plane shear loading. The numerical results partly reflect the expectation that cracks propagate preferentially along the weakest material direction(s). In some cases, propagation is partially seen to occur macroscopically along strong directions, whereas microscopic zig-zagging patterns (where 'microscopic' refers to a size scale of the order of the size of a finite element) are observed. Also, the results suggest an even higher sensitivity of the crack pattern to slight perturbations than already known for the isotropic fracture case \cite{TGstoch2020}.

Future work may include the following:
\begin{itemize}
    \item The proposed formulation for four-fold anisotropy should be numerically tested on more general loading conditions and possibly extended to the three-dimensional case. Also, its results should be thoroughly compared with those of the higher-order models for four-fold anisotropy available in the literature, such as the one in \cite{Li2019};
    \item The stochastic approach advocated in \cite{TGstoch2020} to characterize non-uniqueness of the solution can be applied to the proposed anisotropic \texttt{Foc-}4 formulation, which seems to be even more sensitive to slight perturbations of the system than the formulations for isotropic fracture;
    \item While this study concentrated on two-fold and four-fold anisotropy, other types of anisotropy, possibly featured in new artificial (e.g. 3D printed) materials can be in principle investigated with the same methodology. 
 % see the two simplest examples in Figure \ref{Fig:10}.
%     \begin{figure}[!ht]
%    \begin{center}
%   \includegraphics[width=1.0\textwidth]{gamma1and3.png}
%  \end{center}
%    \caption{The plots of $\gamma_1$ and $\gamma_3$ representing respectively one- and three-fold symmetric anisotropic fracture toughness.}
%    \label{Fig:10}
%    \end{figure}
%\item finally, reliable (that is, robust and accurate) and efficient computational schemes must yet be developed.
\end{itemize}

%%%%%%%%%%%%%%%%%%%%%%%%%%%%%%%%%%%%%%%%%%%%%%%%%%%%%%%%
\section*{Acknowledgements}
The support of the German Research Foundation (DFG) through the project LO 2370/4 `Phase-field computation of brittle fracture: robustness, efficiency, and characterisation of solution non-uniqueness' is gratefully acknowledged. Thanks are also extended to Prof. Gilles Francfort for providing comments on the findings in Appendix \ref{Ap1}.

%%%%%%%%%%%%%%%%%%%%%%%%%%%%%%%%%%%%%%%%%%%%%%%%%%%%%%%%
\appendix

%%%%%%%%%%%%
\section{Existence and uniqueness of the phase-field solution for fixed displacement}
\label{Ap1}
In this appendix, we focus on the isotropic \texttt{Foc-}4 formulation in (\ref{Foc4_iso}) as well as on the anisotropic \texttt{Foc-}2 and \texttt{Foc-}4 formulations in (\ref{Foc2_aniso}) and (\ref{Foc4_aniso}). In all cases, $\mathcal{E}$ is convex with respect to $\bm u$ for fixed $\alpha$, hence the problem $\min_{\bm u}\mathcal{E}$ for fixed $\alpha$ is guaranteed to have a unique solution. Below, we investigate the well-posedness of the problem $\min_{\alpha}\mathcal{E}$ for fixed $\bm u$. To this end, we use the results provided by the direct methods (DM) in the calculus of variations \cite[Theorem 3.3]{Dacorogna2004}: 

{\em
Let $\Omega\subset\mathbb{R}^n$ be a bounded open set with Lipschitz boundary. Let $f\in C^0(\overline{\Omega}\times\mathbb{R}\times\mathbb{R}^n), f=f(x,u,\xi)$ satisfy

(H1)\quad $\xi\rightarrow f(x,u,\xi)$ is convex for every $(x,u)\in\overline{\Omega}\times\mathbb{R}$;

(H2)\quad there exist $p>q\geq1$ and $c_1>0$, $c_2,c_3\in\mathbb{R}$ such that
\begin{equation*}
f(x,u,\xi)\geq c_1|\xi|^p+c_2|u|^q+c_3, \; \forall\; (x,u,\xi)\in\overline{\Omega}\times\mathbb{R}\times\mathbb{R}^n.
\end{equation*}

Let
\begin{equation*}
(P)\quad \inf
\left\{
I(u)=\int_\Omega f(x,u(x),\nabla u(x))\,dx:\;u\in u_0+W^{1,p}_0(\Omega)
\right\}=m,
\end{equation*}
where $u_0\in W^{1,p}$ with $I(u_0)<\infty$. Then there exists $\overline{u}\in u_0+W^{1,p}_0(\Omega)$ a minimizer of (P).

Furthermore, if $(u,\xi)\rightarrow f(x,u,\xi)$ is strictly convex for every $x\in\overline{\Omega}$, then the minimizer is unique.
}

Our findings are summarized in Table \ref{ExUn}, whereas the derivations are outlined in the following.

\begin{table}[ht]
\caption{Existence and uniqueness of the phase-field solution for fixed displacement for the isotropic \texttt{Foc-}$4$ and the anisotropic \texttt{Foc-}$2$ and \texttt{Foc-}$4$ formulations.}
\centering
      \begin{tabular}{c||c|c}
      \hline \rule{0pt}{10pt}
        Model  
        &  Existence  
        &  Uniqueness \vspace{0.1cm} \\
	 \hline\hline \rule{0pt}{14pt}
	    Isotropic \texttt{Foc-}$4$  
	    & {\faThumbsOUp} %\leftthumbsup  
	    & \begin{tabular}{c}
            Cannot be shown \\ by DM
           \end{tabular} \vspace{0.1cm} \\ 
    \hline\hline \rule{0pt}{20pt}
        Anisotropic \texttt{Foc-}$2$  
        & \begin{tabular}{c} {\faThumbsOUp} \\ 
          $\forall\tau\in[0,1),\; \omega\in[0,\pi)$ 
          \end{tabular}  
        & \begin{tabular}{c} {\faThumbsOUp} \\ 
          $\forall\tau\in[0,1),\; \omega\in[0,\pi)$ 
          \end{tabular} \vspace{0.1cm} \\
     \hline \rule{0pt}{30pt}
        Anisotropic \texttt{Foc-}$4$  
        & \begin{tabular}{l} 
          \textbullet\; \faThumbsOUp \\
        \hspace{0.8cm}  if $\tau\in(0,\frac{1}{3}],\;\omega\in[0,\frac{\pi}{2})$ \vspace{0.1cm} \\
%          \hline
          \textbullet\; Cannot be shown by DM \\ 
          \hspace{0.8cm} if $\tau\in(\frac{1}{3},1),\;\omega\in[0,\frac{\pi}{2})$
          \end{tabular}
        &  \begin{tabular}{c}
            Cannot be shown \\ by DM
           \end{tabular}  \vspace{0.1cm} \\ 
	\hline
      \end{tabular}
\label{ExUn}
\end{table}

%%%%%
\subsection{Anisotropic \texttt{Foc-}2 model}
With $\bm u$ fixed, the integrand of the anisotropic \texttt{Foc-}2 model in (\ref{Foc2_aniso}) in the $\alpha$-direction reads:
\begin{equation*}
f({\bf x},\alpha,\nabla\alpha):=    
{\sf g}(\alpha)\Psi(\bm\varepsilon(\bm u))
+\frac{G_0}{2}
\left\{ 
\frac{\alpha^2}{\ell}
+\ell|\nabla\alpha|^2
\right.
\end{equation*}
\begin{equation}
\left.
+\ell\tau
\left[ 
\cos(2\omega)
  \left(\alpha_y^2-\alpha_x^2\right)
-2\sin(2\omega)\alpha_x\alpha_y
\right]
\rule{0pt}{0.5cm}\right\},
\label{f_Foc2_aniso}
\end{equation}
where $\tau\in[0,1)$, $\omega\in[0,\pi)$ and ${\sf g}(\alpha)=(1-\alpha)^2$. 

In the following, we first show that $f$ satisfies a coercivity condition, and then prove that $\bm\xi\rightarrow f({\bf x},\alpha,\bm\xi)$ is convex for every $({\bf x},\alpha)\in\Omega\times\mathbb{R}$, and $(\alpha,\bm\xi)\rightarrow f({\bf x},\alpha,\bm\xi)$ is strictly convex for every ${\bf x}\in\Omega$. The first two properties are required for the existence, and the last one for the uniqueness of the solution.

The coercivity of $f$ is straightforward: using that ${\sf g}(\alpha)\Psi(\bm\varepsilon(\bm u))\geq0$ for all $({\bf x},\alpha)\in\Omega\times\mathbb{R}$ and the first equation in (\ref{infsup_2}) we have
\begin{equation}
f({\bf x},\alpha,\bm\xi)\geq
\frac{G_0}{2}
\left\{ 
\frac{\alpha^2}{\ell}
+(1-\tau)\ell|\bm\xi|^2
\right\}, \quad \forall ({\bf x},\alpha,\bm\xi)\in\Omega\times\mathbb{R}\times\mathbb{R}^2,
\; \tau\in[0,1).
\label{coerc_2}
\end{equation}

To show the convexity of $\bm\xi\rightarrow f({\bf x},\alpha,\bm\xi)$, we analyze the eigenvalues $\lambda_i$, $i=1,2$ of the corresponding Hessian. These read
\begin{equation}
\lambda_{1,2}=G_0\ell(1\pm\tau).
\label{conv_2a}
\end{equation}
and are such that $\lambda_1\geq\lambda_2>0$ for every $\tau\in[0,1)$. (In fact, since $\lambda_2>0$, we have strict convexity of $f$.) Combining this result with (\ref{coerc_2}), existence of the solution to $\min_{\alpha}\mathcal{E}$ follows by Theorem 3.3 in \cite{Dacorogna2004}.

By Theorem 3.3, uniqueness of the solution to $\min_{\alpha}\mathcal{E}$ also follows due to the strict convexity of $(\alpha,\bm\xi)\rightarrow f({\bf x},\alpha,\bm\xi)$. Indeed, the first eigenvalue of the Hessian in this case reads $\lambda_1={\sf g}^{\prime\prime}(\alpha)\Psi(\bm\varepsilon(\bm u))+\frac{G_0}{\ell}$. It is strictly positive for every $({\bf x},\alpha)\in\Omega\times\mathbb{R}$, since ${\sf g}^{\prime\prime}(\alpha)=2$ and $\Psi(\bm\varepsilon(\bm u))\geq0$ in $\Omega$. The other two eigenvalues $\lambda_{2,3}$ coincide with those given by (\ref{conv_2a}).

The above results are summarized in Table \ref{ExUn}. Notice that there is {\em no correlation} of this result with the convexity result for $\gamma_2$ in (\ref{convexity2}).

%%%%%
\subsection{Isotropic and anisotropic \texttt{Foc-}4 models}
Similarly to the previous section, assuming $\bm u$ to be fixed, we consider the integrand of the anisotropic \texttt{Foc-}4 model in (\ref{Foc4_aniso}) in the $\alpha$-direction, namely,
\begin{equation*}
f({\bf x},\alpha,\nabla\alpha):=    
{\sf g}(\alpha)\Psi(\bm\varepsilon(\bm u))
+\frac{G_0}{4}
\left\{ 
\frac{3}{b_\mathsf{w}}\frac{\alpha^4}{\ell}
+\ell^3|\nabla\alpha|^4
\right.
\end{equation*}
\begin{equation}
\left.
+\ell^3\tau
\left[ 
\cos(4\omega)
  \left(
  \alpha_x^4-6\alpha_x^2\alpha_y^2+\alpha_y^4
  \right)
+4\sin(4\omega)
  \left(
  \alpha_x^3\alpha_y-\alpha_x\alpha_y^3
  \right)
\right]
\rule{0pt}{0.5cm}\right\},
\label{f_Foc4_aniso}
\end{equation}
where $\tau\in[0,1)$, $\omega\in[0,\frac{\pi}{2})$, ${\sf g}(\alpha)=(1-\alpha^4)^2$. Since the isotropic model in (\ref{Foc4_iso}) is a limiting case of (\ref{f_Foc4_aniso}) with $\tau=0$, the results apply to this case as well.

The coercivity result is obtained using the first equation in (\ref{infsup_4}) and reads:
\begin{equation}
f({\bf x},\alpha,\bm\xi)\geq
\frac{G_0}{4}
\left\{ 
\frac{3}{b_\mathsf{w}}\frac{\alpha^4}{\ell}
+(1-\tau)\ell^3|\nabla\alpha|^4
\right\}, \quad \forall ({\bf x},\alpha,\bm\xi)\in\Omega\times\mathbb{R}\times\mathbb{R}^2,
\; \tau\in[0,1).
\label{coerc_4}
\end{equation}

For assessing the convexity of $\bm\xi\rightarrow f({\bf x},\alpha,\bm\xi)$, we compute the eigenvalues $\lambda_i$, $i=1,2$ of the corresponding Hessian:
\begin{equation*}
\lambda_{1,2}=G_0\ell^3
\left\{\rule{0pt}{0.4cm}
2(\xi_1^2+\xi_2^2)\pm
\left[   
9(\xi_1^2+\xi_2^2)^2\tau^2
\right.
\right.
\end{equation*}
\begin{equation*}
+6\left(
\cos(4\omega)(\xi_1^4-6\xi_1^2\xi_2^2+\xi_2^4)
+4\sin(4\omega)\xi_1\xi_2(\xi_1^2-\xi_2^2)
\right)\tau
\end{equation*}
\begin{equation}
+\left.\left.
(\xi_1^2+\xi_2^2)^2
\right]^\frac{1}{2}
\right\}.
\label{conv_4a}
\end{equation}
Here, the signs $+$ and $-$ are assumed to correspond to $\lambda_1$ and $\lambda_2$, respectively. Owing to the estimates in (\ref{infsup_4}), for every $\omega\in[0,\frac{\pi}{2})$, the following holds
\begin{equation*}
\lambda_1>\lambda_2\geq
G_0\ell^3(\xi_1^2+\xi_2^2)(1-3\tau).
\end{equation*}
As a result, $\lambda_2\geq0$ iff $\tau\in[0,\frac{1}{3}]$. Using this along with (\ref{coerc_4}), the existence of the solution to $\min_{\alpha}\mathcal{E}$ can be concluded for the isotropic \texttt{Foc-}4 model ($\tau=0$) and for the anisotropic \texttt{Foc-}4 model when $\tau\in(0,\frac{1}{3}]$. Outside this range, the direct methods of the calculus of variations are insufficient to prove the existence result, and one may try to indirectly show existence by using, e.g., the Galerkin approximation \cite{GFcomm2019}. In fact, this is what we do in Section \ref{Res2-anis4} while performing numerical experiments for $\tau=0.5$, a value which lies outside the presumable solution existence range of $[0,\frac{1}{3}]$.

Finally, let us check the strict convexity of $(\alpha,\bm\xi)\rightarrow f({\bf x},\alpha,\bm\xi)$ when $\tau\in[0,\frac{1}{3}]$, required to establish the solution uniqueness in this range. The first eigenvalue of the corresponding Hessian reads
\begin{equation*}
\lambda_1={\sf g}^{\prime\prime}(\alpha)\Psi(\bm\varepsilon(\bm u))+\frac{9G_0}{b_{\sf w}\ell}\alpha^2,
\end{equation*}
with ${\sf g}^{\prime\prime}(\alpha)=56\alpha^6-24\alpha^2$, and the other two eigenvalues $\lambda_{2,3}$ coincide with those given by (\ref{conv_4a}). The positivity of $\lambda_1$ in $({\bf x},\alpha)\in\Omega\times\mathbb{R}$ cannot be proven in general\footnote{It depends on the interplay of $\Psi(\bm\varepsilon(\bm u))$ and the fraction $9G_0/(b_{\sf w}\ell)$, in other words, it is problem dependent.}. As a result, using the direct methods of the calculus of variations, we cannot conclude that $\min_{\alpha}\mathcal{E}$ possesses a unique solution even for the solution existence range of $\tau\in[0,\frac{1}{3}]$.

The above results are summarized in Table \ref{ExUn}. Once again there is {\em no correlation} of this result with the convexity result for $\gamma_4$ in (\ref{convexity4}).

\end{document}